\documentclass[fleqn,useAMS,usenatbib]{mnras}
\usepackage{epsfig}
\usepackage{graphicx}
\usepackage{pdflscape}
\usepackage{textcomp}
\usepackage{float}
\usepackage{comment} 
\usepackage[caption = false]{subfig}
\usepackage{xcolor}

\newcommand{\ltsima} {$\; \buildrel < \over \sim \;$}
\newcommand{\gtsima} {$\; \buildrel > \over \sim \;$}
\newcommand{\lta} {\lower.5ex\hbox{\ltsima}}
\newcommand{\gta} {\lower.5ex\hbox{\gtsima}}
\newcommand{\Epk} {E_{\rm pk}}

\title[]{Non-dissipative photospheres in GRBs: Spectral appearance in the {\it Fermi}/GBM catalogue}
\author[Z. Acuner et al.]{Zeynep Acuner$^{1}$\thanks{email: acuner@kth.se},   
Felix Ryde$^{1}$, \& Hoi-Fung Yu$^{1,2}$\\
\\
$^{1}$Department of Physics, KTH Royal Institute of Technology, and 
the Oskar Klein Centre for Cosmoparticle Physics, \\
AlbaNova, SE-106 91 Stockholm, Sweden\\ 
$^{2}$Faculty of Science, The University of Hong Kong, Pokfulam, Hong Kong
}

\hypersetup{draft}
\begin{document}

\date{Accepted... Received...; in original form ...}

\pagerange{\pageref{firstpage}--\pageref{lastpage}} \pubyear{2019}

\maketitle

\label{firstpage}

\begin{abstract}
A large fraction of gamma-ray burst (GRB) spectra are very hard below the peak. Indeed, the observed distribution of sub-peak power-law indices, $\alpha$, has been used as an argument for a photospheric origin of GRB spectra.  Here, we investigate what fraction of GRBs have spectra that are consistent with emission from a photopshere in a non-dissipative outflow. This is the simplest possible photospheric emission scenario. We  create synthetic spectra, with a range of peak energies, by folding the theoretical predictions through the detector response of the {\it FERMI}/GBM detector. These simulated spectral data are fit with typically employed  empirical models.  We find that the low-energy photon indices obtain values ranging $-0.4 < \alpha < 0.0$, peaking at around $-0.1$, thus covering a non-negligible fraction of observed values. 
These values are significantly softer than the asymptotic value of the theoretical spectrum of $\alpha \sim 0.4$. The reason for the $\alpha$-values to be much softer than expected, is 
the limitation of the empirical functions to capture the true curvature of the theoretical spectrum. 
We conclude that more than a $1/4$ of the bursts in the GBM catalogue have at least one time-resolved spectrum, whose $\alpha$-values are consistent with a non-dissipative outflow, releasing its thermal energy at the photosphere. 
The fraction of spectra consistent with emission from the photosphere will increase even more if dissipation of kinetic energy in the flow occurs below the photosphere.

\end{abstract}


\section{Introduction}  
\label{sec:intro}
It has been argued that the observed characteristics of gamma-ray bursts (GRBs) can be explained by emission from the photosphere in a relativistic outflow. Two possible scenarios have been suggested. First,  the photospheric emission could be accompanied by a non-thermal component \citep[e.g., ][]{Meszaros2002,Ryde2005,Axelsson2012,Guiriec2013, Giannios2019} in a hybrid scenario producing the range of spectral shapes observed. Prominent examples of such cases are GRB090902B \citep{Ryde2010} and GRB190114C \citep[][Fermi collaboration, in prep.]{Wang2019}. Alternatively, the entire emission could be from the photosphere.  In such a case,  dissipation of the flow kinetic energy close to the photosphere is required in order to broaden the spectra to the observed shapes \citep{Rees&Meszaros2005, Peer2006, Giannios2006,  Beloborodov2010, Ryde2010, Ryde2011, Vurm2013}, with examples analysed in \citet{ahlgren2015confronting, VianelloGill2018, Ahlgren2019}. 

However, the simplest scenario for photospheric emission is a non-dissipative outflow (NDP: non-dissipative photosphere, henceforth) in which the thermal energy content of the flow is released at the photosphere, unaltered by heating in  the flow \citep{Goodman1986,Paczynski1986,Rees1994,Ryde2004, Ghirlanda2013, Larsson2015}. In order for such emission to be detectable the flow has to become transparent close to, or below, where the flow saturates to its final outflow Lorentz factor, $\Gamma$. Otherwise, adiabatic expansion will diminish the thermal component and very little emission will be released. For non-dissipative photospheres the expected spectra are known in detail. Most importantly, geometrical effects cause the spectra to differ from a Planck spectrum.  In particular, in the case of the photosphere occurring in the coasting phase, i.e., above the saturation radius, $r_{\rm s }$, the spectrum is significantly different from a Planck function \citep{Goodman1986, Beloborodov2011, Lundman2013}. Its spectral shape is  much broader, while the asymptotic, low-energy spectral slope is still very hard, with $\alpha \sim 0.4$.

The most common way to assess different emission models in GRBs is by studying the sub-peak spectral shape. This is characterised by the photon index, $\alpha$, of the Band function or, similarly, of the cut-off powerlaw function \citep{Band1993}. The values of $\alpha$  can vary significantly depending on the emission process that is taking place. In a optically-thick, thermal scenario, the Rayleigh-Jeans limit has $\alpha = 1$, the Wien limit has $\alpha = 2$, and the non-dissipative photosphere in the coasting phase has the expected $\alpha$ of 0.4. Non-thermal emission is always in the regime $\alpha$ $\textless$ 0.  This limiting value can be reached by synchrotron emission in many specific cases, such as from electrons with a small pitch angle distribution \citep{Lloyd2000},  jitter radiation \citep{Medvedev2000}, with attenuation by scattering \citep{DermerBottcher2000}, and thermal synchrotron emission \citep{Petrosian1981}. Likewise the limiting values of $\alpha = 0$ can be reached by synchrotron self-Compton emission from  monoenergetic electrons scattering off a self-absorbed seed photons field \citep{SternPoutanen2004}.
 More generally, instantaneous or fast synchrotron cooling can explain $\alpha$ $\textless$ -3/2 and slow synchrotron cooling is expected to occur at $\alpha$ $\textless$ -2/3 \citep[e.g.,][]{Tavani1996}.
 
However, before comparing predictions of physical emission models to the observed characteristics of bursts, such as the distibution of $\alpha$-values, limitations of the detector and the analysis methods must also be taken into account. Such limitations could cause spuriously deviation from the expected $\alpha$ \citep[e.g., ][]{Preece1998, Lloyd2000}.  One such limitation is due to the limited band-width of the detector which prevents the full spectrum to be detected. Another limitation is due to the  typically empirical models that are used to fit the data; the Band function and the cut-off powerlaw function \citep[see, e.g., ][]{Yu2016}. If these empirical models do not match the true spectral shape, such as its curvature, then the fitted parameters might not be readily interpreted. 

There are two possible routes to address these limitations. One way is to fit physical models directly to the data, which eliminates the need for fits to  empirical  functions \citep[e.g.][]{Lloyd2000, Zhang&Yan2011, ahlgren2015confronting, VianelloGill2018, Burgess2018, Ahlgren2019, Oganesyan2019}. However, such analysis is computationally very demanding and, moreover, with the present, limited understanding of GRBs it is not clear what models should be used. Alternatively, one can assume a physical model and study what the response of the detector would be to such a model.
This can be done by first producing synthetic data for a particular physical model, by taking into account the limitations of the detector (energy range, photon detection characteristics, etc.). The synthetic data can then be fitted with empirical models, accounting for the limitations of the typically adopted analysis methods. Such a procedure would allow to identify which parameter space of the empirical fit-model that the physical model corresponds to \citep[see e.g., ][]{Burgess2015_alpha}.  Observed bursts which have spectral properties that coincide with the determined parameter space, can therefore be claimed to be compatible with that particular model.
The advantage of this strategy is that the spectral analyses in all previous GRB catalogues can be directly assessed, since these have been produced using empirical models. 

In this paper, we follow the second strategy and study a very specific and restrictive model, namely the non-dissipative photospheric emission (NDP), by producing synthetic GBM observations and by investigating the results of fitted empirical functions.

\section{Properties of non-dissipative photosphere emission}

The simplest photospheric emission spectrum that can be expected from a GRB
is created in an undisturbed outflow, without any significant energy dissipation in the flow. 
The reason for this is that any dissipation of the flow kinetic energy will energise the electron population. Below the photosphere, these electrons subsequently might be able to distort the thermal photon spectrum into broader and more complicated shapes \citep[e.g.,][]{Peer2006, Giannios2006, Vurm2011, ahlgren2015confronting, Chhotray2018}. Numerical simulations of a jet emerging from a progenitor star do indicate that dissipation is expected to some degree \citep[e.g., ][]{Ito2013, DeColle2017}. However, the scenario without dissipation is still of interest to study, since such spectra define the narrowest and thereby the most extreme spectral shapes expected in the photospheric scenario. Moreover, if such spectra are identified in real cases, a strong limit on the degree of dissipation can be put.


But even in the absence of dissipation, there are still factors that cause the spectrum to differ from the comoving and local original thermal shape. One such factor is the fact that the photons' last scattering off electrons in the flow occur at significantly different radii  \citep[e.g.,][]{Peer2008, Beloborodov2011}. This could lead to a broader spectrum, if the temperature of the flow varies with radius. Another factor is the angular distribution of the photons in the lab frame, that is  affected by the radial expansion; their distribution becomes more and more  anisotropic the closer they are to the photosphere \citep{Beloborodov2010}. 
Finally, differences in the magnitude of the Doppler boosts for emission at high latitudes will also cause a broadening \citep{Abramowicz1991, Peer2008}. 

\begin{figure}
 \includegraphics[width=\columnwidth]{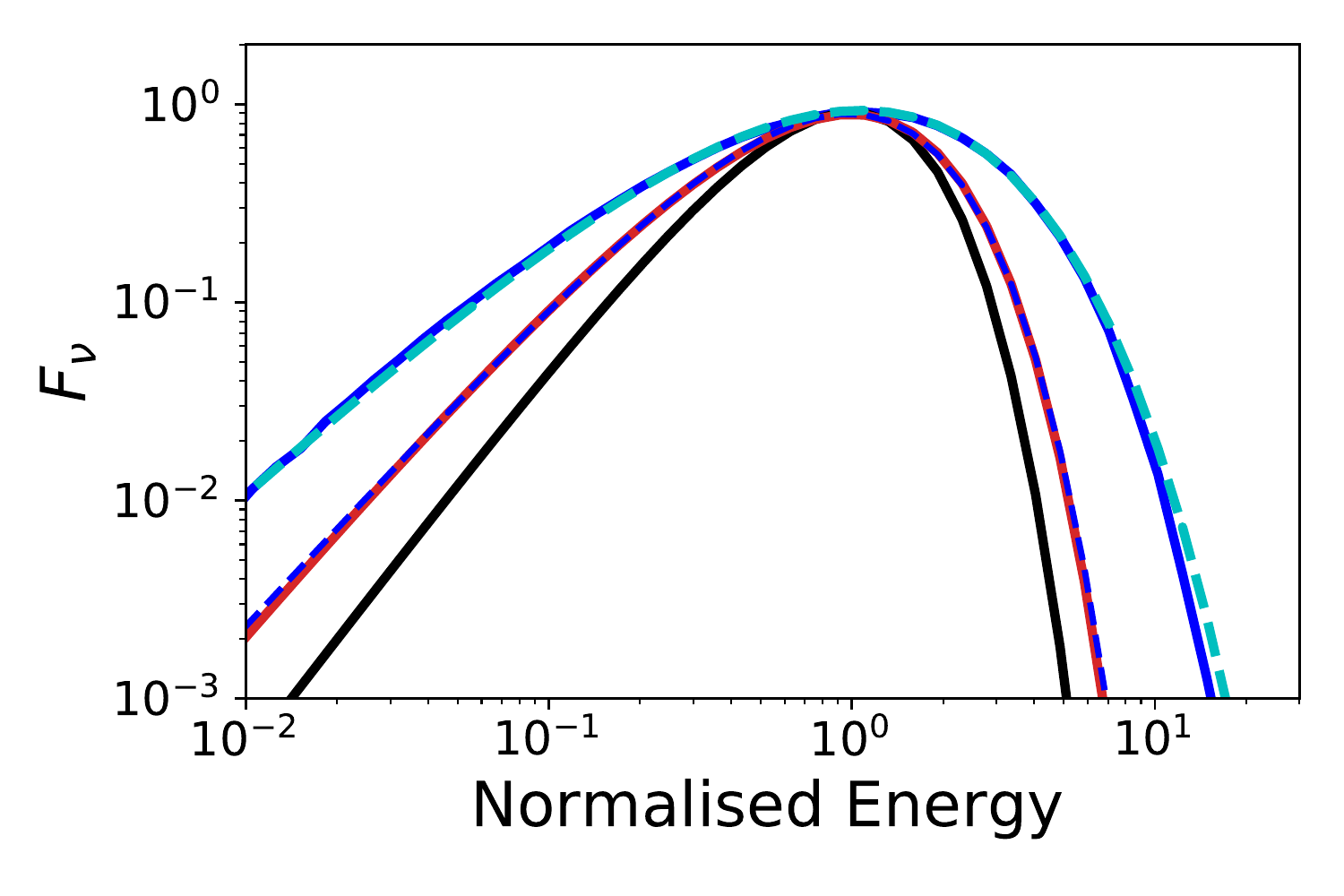}
 \caption{Energy spectra ($F_\nu$; arbitrary units) from non-dissipative photospheres (NDP). The blue lines are for a photosphere occurring the coasting phase and the red line is for an acceleration phase photosphere. The black line is for a Planck function and is shown for comparison. The dashed lines show the approimations given by Eqns. (\ref{eq:1}) and   (\ref{eq:2}).}
\label{fig:1}
\end{figure}

These factors cause the spectrum from a photosphere occurring above the saturation radius (where the Lorentz factor saturates) in a non-dissipative flow to be broader than a Planck spectrum \citep{Goodman1986, Beloborodov2011, Lundman2013}. There is no simple analytical expression for this spectrum, and it has to be derived numerically. However, the shape of its photon flux spectrum can be approximated by a powerlaw with a stretched exponential cut-off:
\begin{equation}
    N_{\rm E} =  K \, \left( \frac{E}{E_{\rm pivot}}\right)^{0.4} e^{-\left(\frac{E}{E_{\rm c}}\right)^{0.65}},
\label{eq:1}
\end{equation}
\noindent 
where $N_{\rm E}$ is the photon flux (unit), $E_{\rm pivot}$ is the pivot energy and $E_{\rm c}$ is the cut-off energy. In Figure \ref{fig:1} the approximation given by Equations (\ref{eq:1}) is shown  as the light blue dashed line, which is overlayed on the numerically calculated spectrum in \citet{Lundman2013} (dark blue, solid line; see Fig. 1 in Ryde et al. 2017). In the figure the black line is the Planck function, aligned to the same spectral peak. The approximation in Eq. (\ref{eq:1}) is useful since it can easily be implemented in spectral analysis tools, such as {\tt RMfit}\footnote{http://fermi.gsfc.nasa.gov/ssc/data/analysis/rmfit/}, {\tt XSPEC}  \citep{Arnaud1996}, and {\tt 3ML} \citep{Vianello2017}. 

If, on the other hand, the photosphere occurs well below $r_{\rm s}$, then these factors will have a smaller effect and the narrowest allowed spectrum will be emitted. An analytical equation for this spectrum is given by Eq. (2) in \citet{Ryde2017}, but can again be approximated by a simple analytical function; a cut-off powerlaw:
\begin{equation}
    N_{\rm E} =  K \, \left( \frac{E}{E_{\rm pivot}}\right)^{0.66} e^{-\left(\frac{E}{E_{\rm c}}\right)}
\label{eq:2}
\end{equation}
\noindent
The approximation in Equation (\ref{eq:2}) is shown in Figure \ref{fig:1} as the black, dashed line which is overlayed on the analytical expression in \citet{Ryde2017} (red, solid line).

\section{Simulation and analysis of spectra}

All the analysis for this study has been carried out within the The Multi-Mission Maximum Likelihood {\tt 3ML} framework  \citep{Vianello2017}. As a first step, we have generated synthetic spectra \textit{observed} by the Gamma-ray Burst Monitor (GBM) onboard the {\it Fermi Gamma-ray Space Telescope}.  GBM consists of 12 Sodium-Iodide (NaI) with $E$ $\epsilon$ ($8 - 900$ keV) and two Bismuth-Germenate (BGO) with $E$ $\epsilon$ ($205$ keV - $40$ MeV) detectors, totaling to 14 detectors. 


To create these \textit{observed} spectra, the raw model should be folded by the detector response, which is taken from the instrument's standard response files that are available for every observation. We have chosen GRB120711115 for this task which is a generic long burst with a non-thermal looking spectrum and a large $\Epk$. The simulated data consists of two NaI (NaI 2 and NaI A) and one BGO (BGO 0) detector that have the smallest source viewing angles in this particular observation, following the procedure of the {\it Fermi} GBM time-integrated and time-resolved catalogues \citep[]{Goldstein2012, Gruber2014, Yu2016}. The source and background intervals are obtained from the GBM catalogue. The source time interval is chosen as 62 s to 106 s after the trigger time (the fluence time). The background post-source interval is between 146-191 s while the pre-source interval is from -45 to -4 s (pre-trigger). The background selections are modeled with a zeroth order polynomial in time which is determined by a likelihood test by fitting the total count rate first. Following this, the polynomial is fitted to all energy channels and integrated over time to estimate the count rate from the background in each channel and their respective errors. It should be noted that the burst choice for this analysis does not effect the results since the fitted spectra are simulated from a theoretical model and only the response files are made use of from the chosen data files.


The synthetic spectra are then fitted within the Bayesian inference framework. All spectra are modeled with a powerlaw with an exponential cut-off. This is an empirical model that has been extensively used in the community, e.g. in the {\it Fermi} GBM catalogues \citep{Goldstein2012, Gruber2014, Yu2016, Yu2019}. For posterior simulations we have picked informative priors that specify realistic parameter intervals that can be detected with GBM and are reasonable for capturing the shape of our seed function. The normalization ($K$) is assigned a log uniform prior with $K$ $\sim$ ($10^{-11}$, $10$) cm$^{-2}$ keV$^{-1}$ s$^{-1}$. The low energy index ($\alpha$) and the cut-off energy ($E_{C}$) are assumed to have uniform priors of $\alpha$ $\sim$ ($-3$ , $2$) and $E_{C}$ $\sim$ ($1$ , $10000$) keV respectively. The posterior sampling was performed via the $\texttt{emcee}$ (an affine invariant Markov Chain Monte Carlo (MCMC) ensemble sampler) implementation in $\texttt{3ML}$ \citep{emcee}. The model viability has been tested through posterior predictive checks and the convergence of MCMC simulations have been checked for, for which the details can be found in Appendices A and B.

We have implemented both Maximum Likelihood Estimate (MLE) and Bayesian methods in the simulation process to be able to double check. We find perfect agreement between the frequentist Maximum Likelihood (MLE) and Bayesian framework results.

\section{Empirical characterisation of NDP photon spectra}

Typically GRB spectra are described by empirical models.  If these empirical models do not capture the true curvature of the  theoretical emission spectrum, then the empirical model parameters might attain spurious values. This is particularly the case when the spectra are studied over a limited energy range. This has, for instance, been shown  for synchrotron emission spectra \citep{Lloyd2000}. They showed that by using an empirical model, such as the Band function, to describe the synchrotron emission spectrum, the determined low-energy photon-index, $\alpha$, will depend on the position of the spectral peak, relative to the energy of the low-energy detector limit. Indeed, the fitted values of $E_{\rm pk}$ and $\alpha$ will give a well-defined correlation, even though the underlying spectral shape is unaltered.  

On the other hand, if the empirical model used is able to properly capture the curvature of the theoretical emission spectrum, then the determined parameters will be valid, even outside of a limited energy range. For instance, \citet{Burgess2015_alpha} showed that if the true spectrum is a Band function, then the correct parameters are retrieved, independent of the position of the spectral peak. Likewise, \citet{Ryde2019} showed that this is also the case for a photospheric spectrum arising during the acceleration phase, when it is fitted by a cut-off powerlaw function. The determined $\alpha$-value is independent of the position of the peak relative to the low-energy detector limit. In such cases the curvature imprinted in the spectrum will be enough to find the correct parameter values, 
even though its asymptotic value has not been reached within the observed energy range. 
 
However, since the true, theoretical emission model is not yet known, such spurious effects must be taken into consideration when interpreting data and their correlations. One way to do this is to analyse fits to the analytical (theoretical) spectra using  empirical functions (\S \ref{sec:curvature}) and another way is to analyse fits to synthetic spectra (\S \ref{sec:synthspec}), that are based on theoretical emission models. Below, we perform such  analyses on the spectrum from a non-dissipative photosphere.


\subsection{Curvature of a non-dissipative photosphere spectrum}
\label{sec:curvature}

We will now investigate how well the empirical functions 
can mathematically characterise the theoretical spectra from a photosphere occurring during the coasting phase in a non-dissipative flow, e.g., given by Eq. (\ref{eq:1}).

\begin{figure}
 \includegraphics[width=\columnwidth]{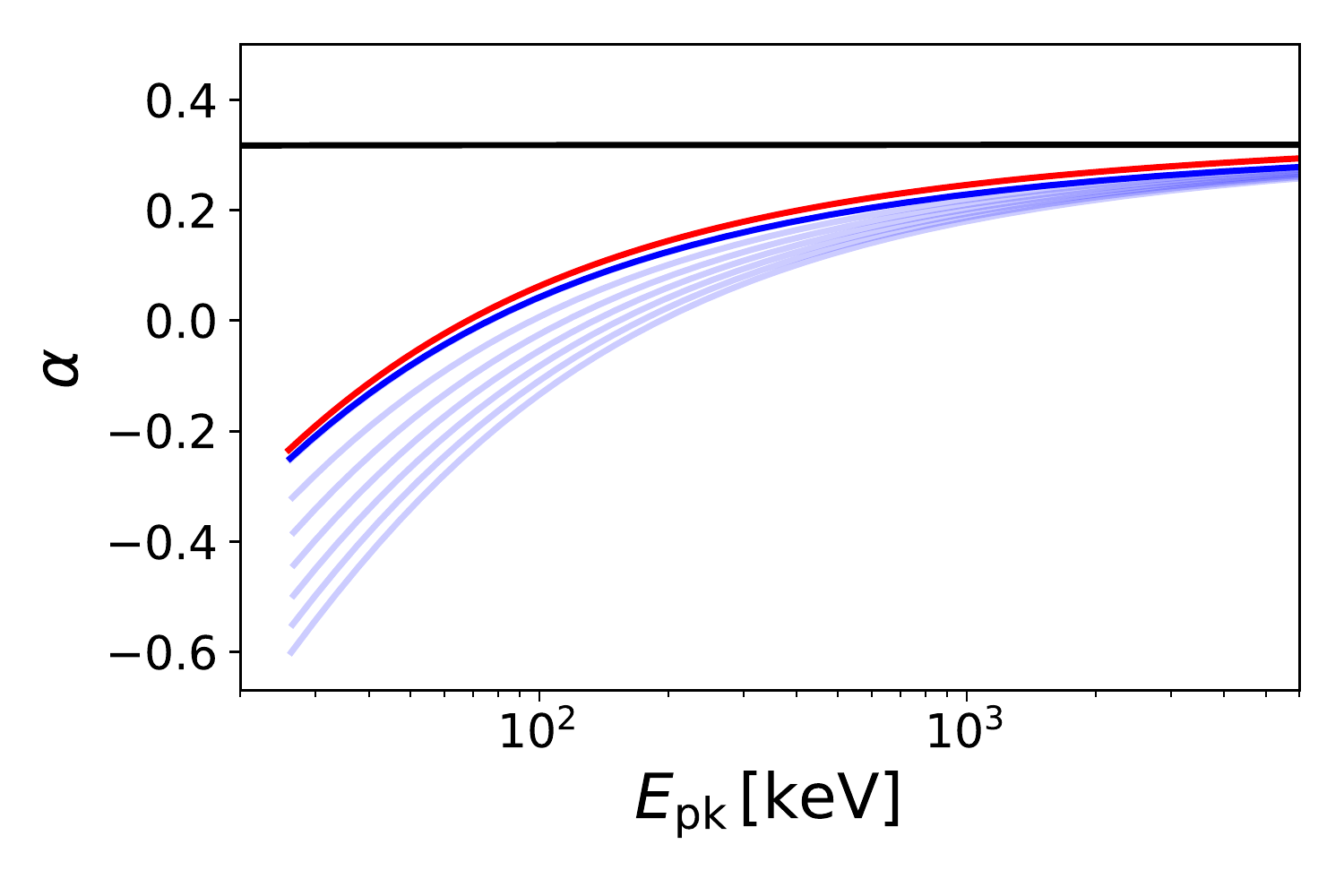}
 \caption{Relation between $\alpha$ and $E_{\rm pk}$ of the cutoff powerlaw function, resulting from fits to a NDP spectrum. The fits are performed over different spectral ranges. 
 Black line: no limits, i.e., all energies are used. Blue lines: A range of  low energy limits are imposed, $E_{\rm low} = [8, 10, 12, 14, 16, 18, 20]$ keV, which yield the blue line relations, from top to bottom, respectively. The red line corresponds to Band function fits with $E_{\rm low} = 8$ keV.
}
 \label{fig:2}
\end{figure}

We compared Eq. (\ref{eq:1}) to a cut-off powerlaw function, over different energy ranges. First, we do this by fitting a cut-off powerlaw function, without limiting the energy range. The value of the fitted powerlaw index $\alpha$ is shown as the black line  in Figure \ref{fig:2}. The value is consistently $\alpha = 0.32$. We note that this value is slightly different from the 
asymptotic powerlaw slope which has a photon index of $0.4$ \citep{Beloborodov2010}, giving a first indication that the cut-off powerlaw has a limited ability to perfectly reproduce the spectral shape. 

Second, we limit the spectral range, over which the fit is performed, to be above $E_{\rm low} = 8$ keV. This corresponds to the low-energy threshold of the GBM detector \citep{Meegan2009_GBM}. We find that the determined $\alpha$-value will depend on the position of the peak energy, $E_{\rm pk}$. The dark blue line in Figure \ref{fig:2} shows this relation between $\alpha$ and $E_{\rm pk}$. The value of $\alpha$ is consistently above 0 for $E_{\rm pk}$ larger than 100 keV. 

A similar study was then done using the Band function as the empirical model. The corresponding result is shown in the red curve in Fig. \ref{fig:2}. The results are very similar to the fits with the cut-off powerlaw function, indicating that the following investigation is valid for both functions.

We then tried a range of different low-energy limits, $E_{\rm low}$. A motivation for this is that the effective area of GBM does not reach its maximal value until 20 keV and therefore photon detections above 20 keV are more significant than below. The resulting correlations are shown by the series of light blue lines, using the range $E_{\rm low} = [10, 12, 14, 16, 18, 20 ]$ keV. In the case of  $E_{\rm low} = 20$ keV, we note that a range of $\alpha > - 0.6$ can be reached.




In summary, by just narrowing the analysed energy range, a positive correlation between fitted $E_{\rm pk}$ and $\alpha$ is expected, as an energy-window bias effect. Significant deviation occurs for $E_{\rm pk}$ below 100 keV. In particular, for spectra with $E_{\rm pk}$ larger than a few hundred keV, $\alpha$ is still very hard, being positive.

\subsection{Synthetic {\it Fermi}/GBM spectra from a photosphere}
\label{sec:synthspec}

While we above studied the functional spectral shape, we will now turn to 
simulating GBM data from the theoretical models. This produces synthetic photon data that can be studied with the typical data analysis methods and tools. These data will include all the effects of the dispersion of the instrument and the effective area variations according to the determined detector responses, which is not included in \S \ref{sec:curvature}.

We start by producing a series of synthetic spectra which all have a similar signal-to-noise ratio, SNR $\sim 20$, but with different peak energies. We choose peak energies in the range 40--2000 keV, equally spaced in logarithmic scale. This is the range over which $E_{\rm pk}$ is typically observed \citep{Yu2016}. 

\begin{figure}
 \includegraphics[width=\columnwidth]{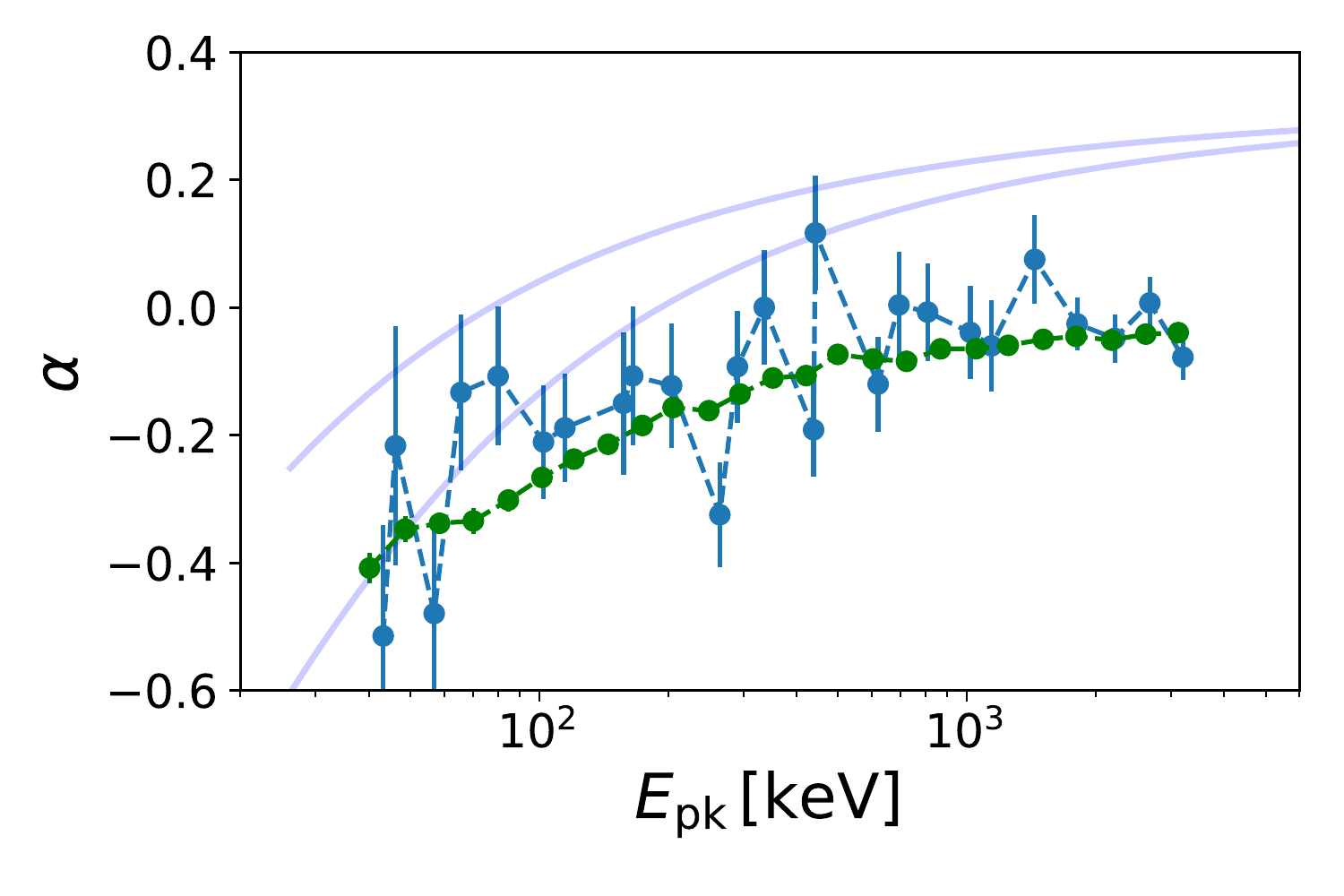}
 \caption{Same as Figure \ref{fig:2}. The data points are from fits of the cutoff powerlaw function to synthetic NDP spectra. The synthetic spectra are simulations yielding what they would have been measured like by the GBM detector. Spectra with twenty-five different $E_{\rm pk}$-values are simulated. The blue points correspond to the fits  of spectra that have SNR=20, while the green points correspond to  SNR=300.}
 \label{fig:data}
\end{figure}

These spectra are then fitted with a cut-off powerlaw.  The results of the fits are shown  as blue data points in Figure \ref{fig:data}. In the figure we also plot the relations for $E_{\rm low} = 8$ keV and $20$ keV from \S \ref{sec:curvature} for comparison. Two features can be noted. First, a positive correlation between the fitted $\alpha$-value and $E_{\rm pk}$-values, is in accordance to the window effect studied above, in \S \ref{sec:curvature}. Second, there is an obvious offset between the mathematical fits over a limited energy band (light-blue lines) and the $\alpha$-values found from folding the spectra through the detector response (blue data points). Instead of the expected value of  $\alpha \sim 0.3$ at large $E_{\rm pk}$-values, the synthetic spectra have fitted values of  $\alpha \sim -0.1$.  

We then produce a new series of synthetic spectra but now with a SNR $\sim 300$, in order to investigate whether the data significance affects the result.  The $\alpha$-values from these fits are depicted by the green data points in Figure  \ref{fig:data}. These values are only slightly more negative  and the conclusions drawn are the same as for the case with SNR $\sim 20$.

\begin{figure}
 \includegraphics[width=\columnwidth]{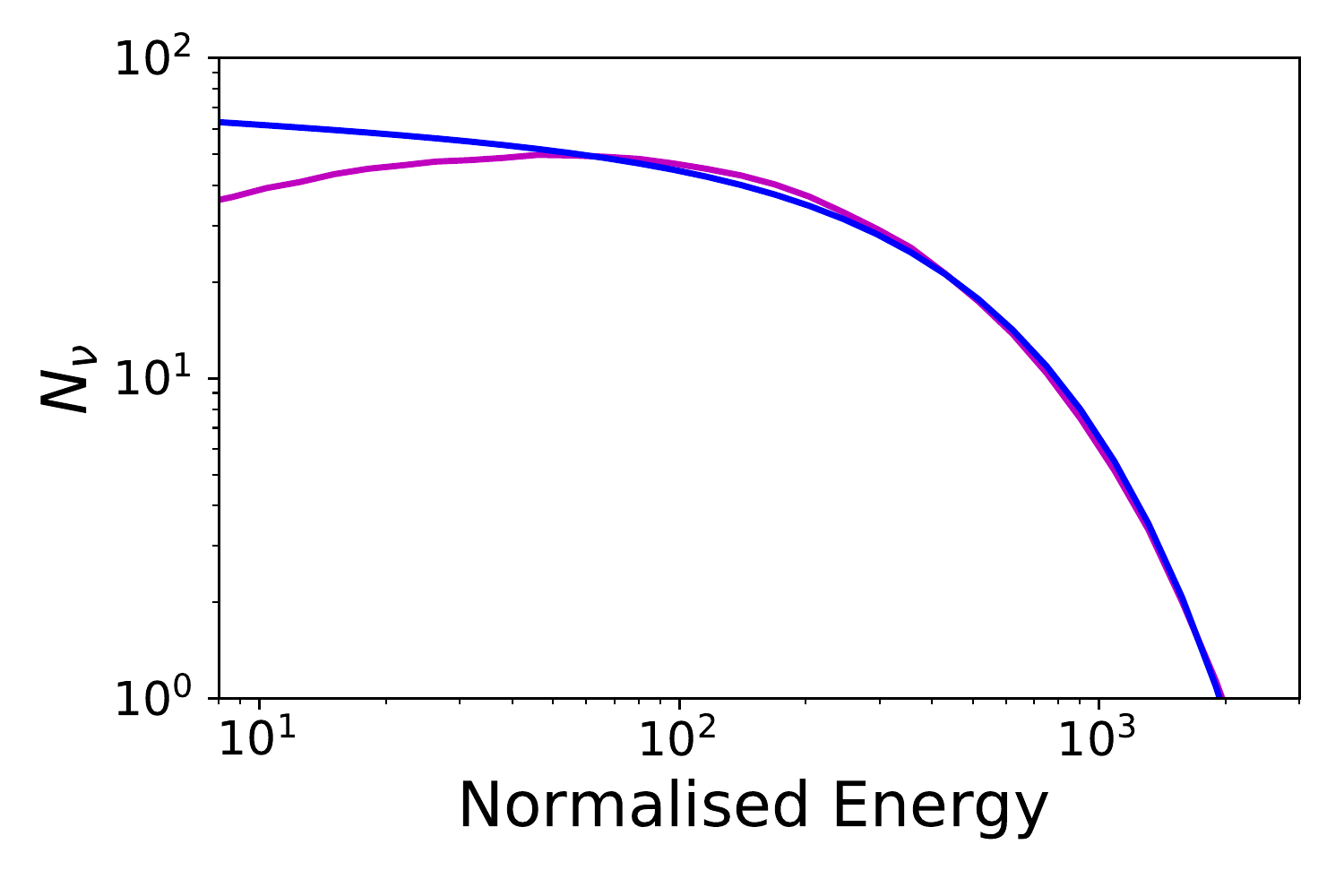}
 \includegraphics[width=\columnwidth]{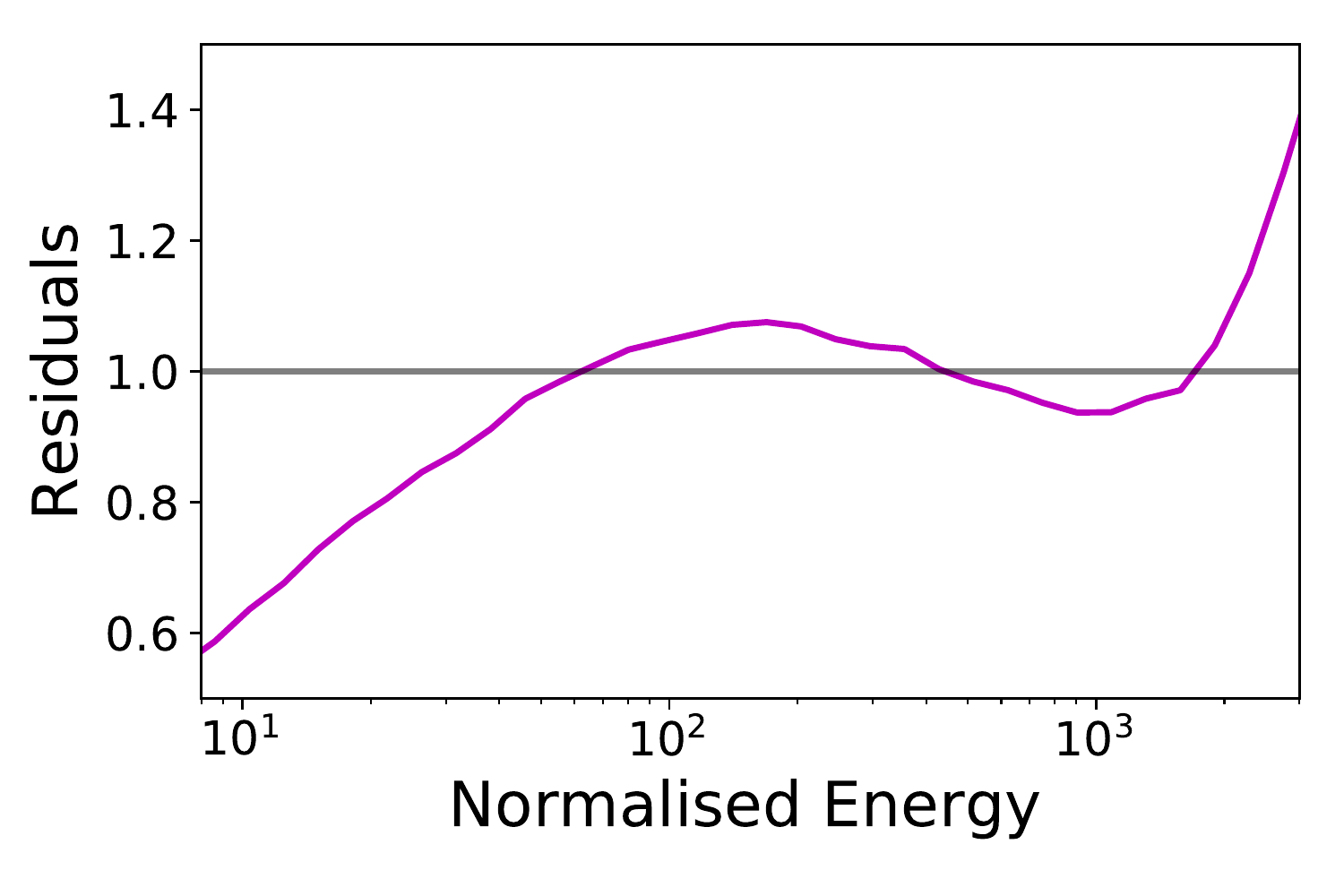}
 \caption{Comparison of between model and fitted photon flux spectra ($N_{\nu}$; arbitrary units). Upper panel:
 Theoretical model spectrum with $E_{\rm pk} = 500$ keV (purple line) and the corresponding best-fit spectrum, using a cutoff powerlaw function (blue line). Lower panel: Residuals between the model and the best-fit function. 
  }
 \label{fig:moda}
\end{figure}

In order to investigate the reason for the off-set found in Figure \ref{fig:data}, we plot in the upper panel in Figure \ref{fig:moda} both the theoretical input spectrum, from which the synthetic data were produced (purple line), and the best-fit to these data by a cutoff powerlaw function (blue line). In this particular simulation $E_{\rm pk} = 500$ keV. It is clear from this figure that the best-fit value of $\alpha$ is much softer than the actual spectral slope below the peak, elucidating the off-set seen in Figure \ref{fig:data}. However, the fit captures the spectral shape relatively well above 50 keV. In the lower panel in Figure \ref{fig:moda}  the residuals of the ratio of the input spectrum (data)  and the best-fit spectrum (model) is shown. 
The fit leaves residuals with a particular, wavy structure.  The residuals are the smallest (within 10 \%) between 35--2000 keV.


In summary, the cutoff powerlaw function does not properly describe the curvature of the spectrum from a non-dissipative photosphere. Therefore, as shown in Figure \ref{fig:data}, the determined  values of $\alpha$, from  such a photosphere observed by GBM, lie around $\alpha \sim -0.1$, significantly differing from the asymptotic powerlaw slope of 0.4. The window effect (positive correlation between $E_{\rm pk}$ and $\alpha$) is mainly affecting the determined values of $\alpha$ below 100 keV. 
Therefore, one should expect $-0.4 < \alpha <0$ from a non-dissipative photosphere observed by GBM.


\section{Catalogue simulations} 
\label{sec:Catalogue}


We will now redo the analysis done in the previous section, but we will instead randomly sample the $E_{\rm pk}$-values from the observed distributions from the GBM catalogue. This will give us the expected distribution of $\alpha$, in the case all observed spectra were to stem from non-dissipative photospheres.\footnote{The energy peak of the coasting NDP corresponding to peak flux $\Epk$ values determined with the cut-off powerlaw function is first sampled randomly from a lognormal distribution with  8 $\le$ $\Epk$ $\le$ 2000 keV. For each of the 100 spectra that are created, the normalization of the function is adjusted so that SNR $\approx$ 30 (also $\approx$ 300 to examine the high SNR condition).}

We will sample from five different distributions. First, we will sample from the full  $E_{\rm pk}$-distribution in \citet[][]{Yu2016} (\S \ref{sec:full}). Second, we will sample from the $E_{\rm pk}$-distributions belonging to the separate burst clusters that are defined in \citet[][]{Acuner2018} (\S \ref{sec:all}). We will focus on their clusters 1, 3, and 5, which all are characterized by having hard spectra.  Finally, we will sample from the $E_{\rm pk}$-distribution associated to the hardest spectra ($\alpha_{\rm max}$) in \citet[][]{Yu2016} (\S \ref{sec:amax}). The graphical results will be presented as Kernel Density Estimation (KDE) curves estimated with normal kernels.\footnote{See https://docs.scipy.org/doc/scipy/reference/generated \newline
/scipy.stats.gaussian\_kde.html} 

\begin{figure}
 \includegraphics[width=\columnwidth]{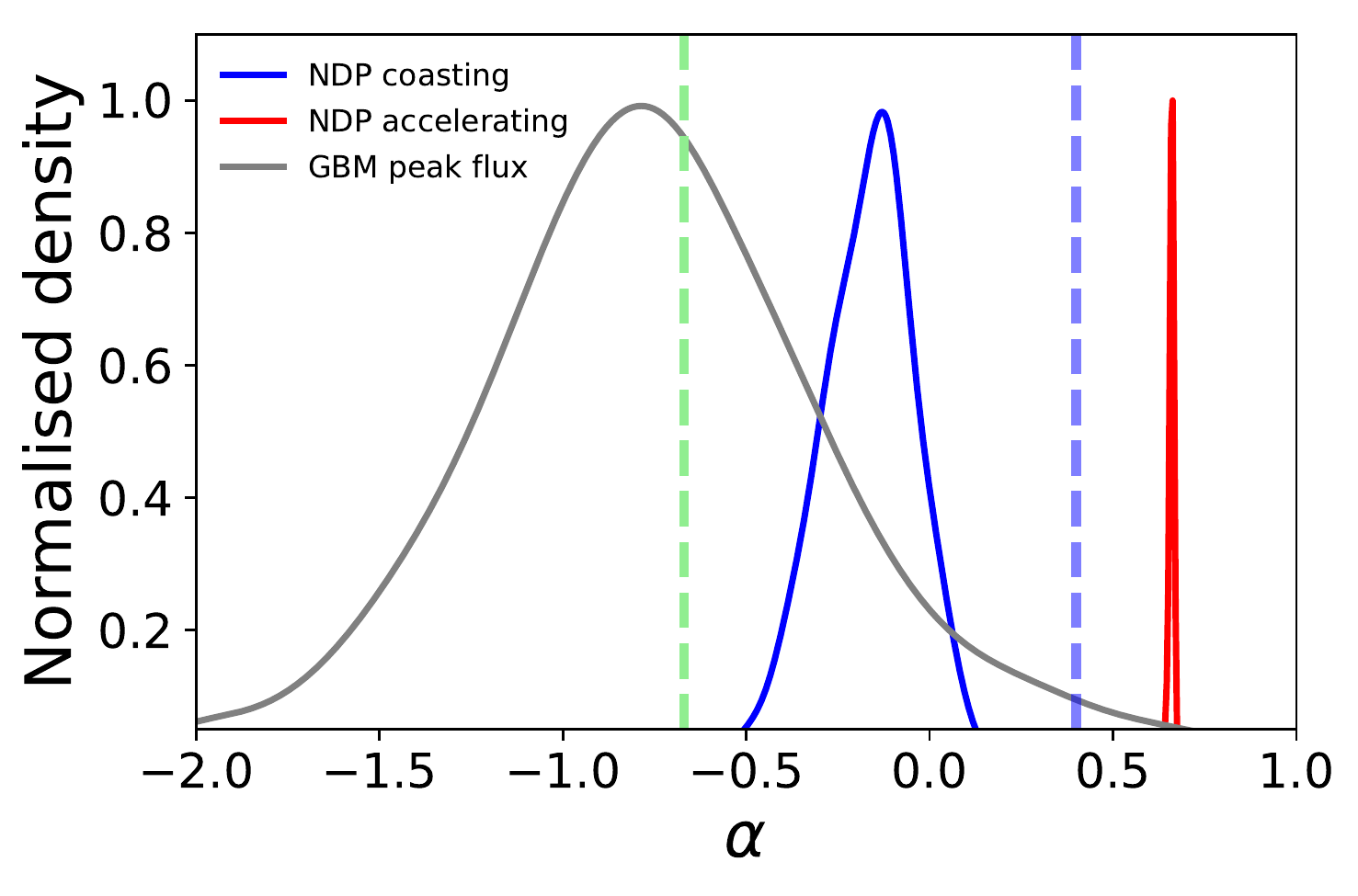}
 \caption{Normalised density distributions of $\alpha$-values  shown as  KDE curves.
 The grey distribution shows the distribution of  $\alpha$-values from the current GBM catalogue (i.e., peak-flux spectra until 19 January 2019, sample size of 2285). The blue distribution is found from fits to synthetic NDP spectra (sample size of 100),  for which the photosphere occurs in the coasting phase, while the red distribution is for photospheres occurring in the accelerating phase (sample size of 100), instead. The dashed green line shows $\alpha=-2/3$, and the dashed blue line respresents the theoretically expected value of the low-energy index for NDP coasting spectra, which is $\alpha=0.4$.
The $E_{\rm pk}$-values, that were used for the simulations of the synthetic spectra, were randomly drawn from their distribution in the GBM catalogue. 
}
\label{fig:all}
\end{figure}

\subsection{Synthetic $\alpha$-distribution by sampling from the GBM catalogue $E_{\rm pk}$-values}
\label{sec:full}

We sample from the $\Epk$-distribution found from fits with the cut-off powerlaw function, made in the {\it Fermi}/GBM catalogue. The selected sample consists of 2228 bursts which have been detected by GBM until 19 January 2019. The median of the distribution is $\sim$ 260~keV, with minimum and maximum values of approximately 10 keV and 10 MeV, respectively.

We have randomly sampled from the above mentioned distribution to assign energy peaks to 100 NDP spectra simulations. These are then fitted with the cut-off powerlaw function. The resulting low-energy index ($\alpha$) distribution is shown by the blue line in Figure \ref{fig:all}. The median value of the distribution is -0.16 and the inter-quantile range (IQR) is 0.15. This distribution should be compared to the asymptotic value of the theoretical NDP spectrum which is $\alpha = 0.4$ (blue dashed line in Fig. \ref{fig:all}). There is a significant offset between the two $\alpha$ estimations.

\begin{table*}
	
    \centering
	\caption{The percentages of simulated NDP spectra that can explain the $\alpha$-values in different subsamples in the the GBM catalogue. 
	}
	\label{tab:1}

    \begin{tabular}{l|ll|l|l|ll}
     Simulation  &   ~     & Sample           & within FWHM (\%) &$\alpha_{\rm min} < \alpha < \alpha_{\rm max}$ (\%) &  $\alpha$ $>$ $\alpha_{\rm min}$ (\%) \\   
     \hline
    NDP coasting phase &    ~              &Full catalogue         & 12  &   28 & 38                                         \\ \hline
    NDP accelerating phase   &     ~     &       Full catalogue          & 0  &   1 & 3                                           \\ \hline
    NDP coasting phase & ~ & Cluster 1  &         14 &   34 & 43                                      \\ 
    ~  & ~ & Cluster 2               &      2  &   2 & 9                                                \\ 
    ~  & ~ & Cluster 3              &         20  &   43 & 54                                             \\ 
    ~  & ~ & Cluster 4               &         0  &   0 & 0                                             \\ 
    ~  & ~ & Cluster 5                &         9  &   30 & 43                                           \\ 

		\hline
	\end{tabular}
\end{table*} 

 Next, we  redo the analysis by creating synthetic data from a photosphere occurring when the flow is still undergoing acceleration (Eq. \ref{eq:2}). This spectrum is much narrower and it can be approximated by an exponential cut-off  powerlaw function. As mentioned above, a consequence of this is that the use of the cutoff powerlaw function in the fitting procedure will not give rise to a window bias effect. Thus, an artificial correlation between $\alpha$ and $E_{\rm pk}$ is not expected. The result from using the distribution of $E_{\rm pk}$ from the full GBM catalogue 
is  shown by the red curve in Figure \ref{fig:all}.  The median value of the distribution is $\alpha_{med} = 0.65$. 


As a comparison the distribution of $\alpha$-values from the fits to the peakflux spectra in the GBM catalogue is shown by the grey line in Figure \ref{fig:all}. For this distribution, we have applied the restrictions that the measured $\alpha$-values should be smaller than 3 and the $\alpha$-errors should be  smaller than 1. This was done in order to remove obvious, and most likely, erroneous fits. 

It should be noted that the peakflux spectra are produced by integrating the emission over approximately 1 second around the flux peak \citep[e.g., ][]{Gruber2014}. This means that they are not necessarily time-resolved, in the sense that they could still contain significant variability (see, e.g., \citet{Golkhou&Butler2014}, \citet{Golkhou&Butler2015}, and \citet{Acuner2018}). 
Bursts with significant spectral evolution, on a time scale much shorter than the integration time scale for these spectra, will thus not reveal the instantaneous emission spectrum. As a 
consequence, the measured $\alpha$-values from the peakflux timebin might be significantly softer than that given by the emission process. 
In any case, the comparison of the $\alpha$-distribution for the  coasting NDP spectra (blue curve) with the catalogue distribution (grey curve) shows that around 12\% to 28\% (see Table \ref{tab:1}) of the catalogue peak flux spectra can be attributed to a non-dissipative photosphere in the coasting phase. This is determined by identifying the percentage of bursts in the GBM distribution that is inside the FWHM of the NDP $\alpha$-distribution for the former and inside the minimum and maximum values for the latter. The fraction of bursts that have $\alpha$-values larger than the minimum value is 38\%. In contrast, only a small fraction of bursts from the GBM samples can be explained by photospheric spectra occurring during the acceleration phase (red curve). 

\subsection{Synthetic $\alpha$-distribution by sampling $E_{\rm pk}$-values from burst clusters}
\label{sec:all}

\begin{figure}
 \includegraphics[width=\columnwidth]{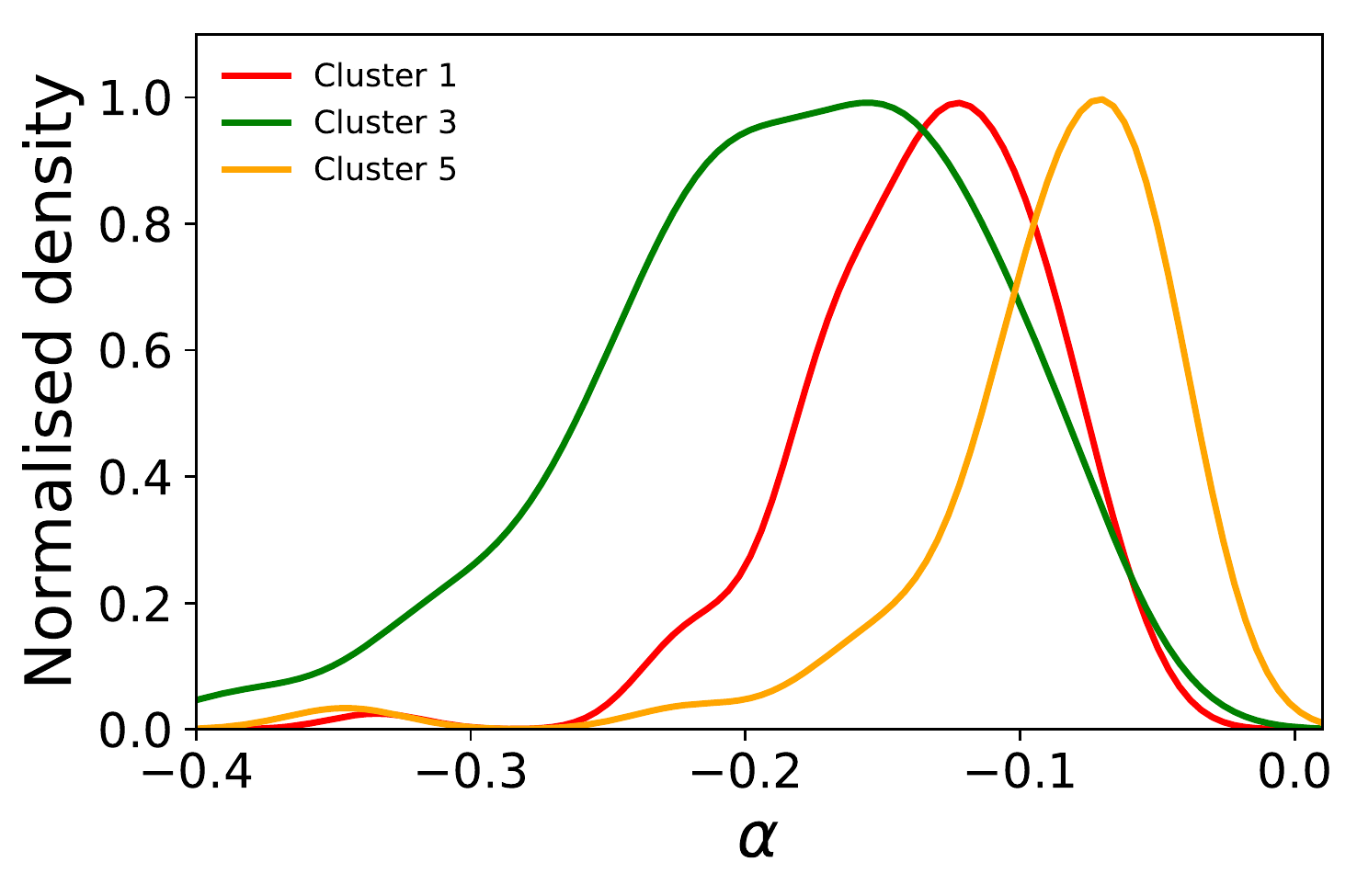}
 \caption{Same as Figure \ref{fig:all}. Here instead the sample of synthetic bursts have $E_{\rm pk}$-values that are randomly drawn from three different $E_{\rm pk}$-distributions. These correspond to clusters 1 (red), 3 (green) and 5 (orange) defined in \citet{Acuner2018} (sample sizes of 100). Medians of the distributions are $\alpha_{\rm 1, \,  med} =  -0.13$, $\alpha_{\rm 3, \,  med} =  -0.18$ and $\alpha_{\rm 5, \,  med} = -0.08$ for clusters 1, 3, and 5, respectively (SNR 30). }
\label{fig:clusters}
\end{figure}

It is often of value to carry out an Exploratory Data Analysis (EDA) to detect any structure that might be innate in the data without assuming any models. This procedure can also reveal unexpected aspects of the data that have not been captured by existing models. \citet[][]{Acuner2018} have carried out such an analysis applying an unsupervised clustering method to the Fermi GBM data which included the Band fit parameters from the GBM catalogue ($\alpha$, $\beta$, $\Epk$) alongside the fluence and $T_{90}$. They identify five clusters and argue that 1/3 of the bursts could be  explained by synchrotron emission (their clusters 2 and 4), whereas 2/3 are explained by a photospheric origin (clusters 1, 3, and 5). The latter class of bursts, which have hard $\alpha$-values, is of particular interest for this study. 

We therefore redid the simulations and analysis made in \S \ref{sec:full}, but now using the $\Epk$ distributions from the five individual clusters from \citet{Acuner2018}, instead. The  basic statistical properties of all the simulated clusters are given in Table \ref{tab:2} and the distributions of clusters 1, 3, and 5 \citep[photosphere clusters in][]{Acuner2018} are shown in Figure \ref{fig:clusters}.
The main difference from the  full catalogue (simulated) distribution (\S \ref{sec:full}; blue line in Fig. 5) is that the IQR are significantly smaller for clusters 1, 3, and 5.
The main source of dispersion in the full catalogue  (simulated) distribution of $\alpha$ is from the bursts in clusters 2 and 4.
The only significant shift in median value is for cluster 5, which has a harder $\alpha$-distribution. This is due to the input $\Epk$ distribution of the cluster 5 simulations. Compared to the other clusters, cluster 5 has the $\Epk$ distribution with the highest $\Epk$ mean \citep[see][]{Acuner2018}.

These distributions can now be compared to the distributions in the original, measured GBM clusters in \citet{Acuner2018}, for which the percentages are presented in their Table \ref{tab:1}.
Furthermore, in Appendix C, we present all five simulated distributions with a comparison to the original data distributions from \citet{Acuner2018}. The fraction of bursts in the five clusters
that can be attributed to a NDP in the coasting phase are also shown in Table \ref{tab:1}.

This comparison indicates that it is mainly clusters 1, 3 and 5 that show some overlap. The overlap for clusters 1 and 5 are still similar to that of the full catalogue, though. However,  cluster 3 has the largest overlap; only less than half of the bursts have  $\alpha$-values that are softer than expected from a NDP. A discussion of a possible cause for this is given in \S \ref{sec:HE}.

On the other hand, clusters 2 and 4 have very small overlap.
This is not surprising since these two clusters are the ones with softest low-energy indices with median values of $\alpha$ are -0.77 and -1.43, respectively \citep[][]{Acuner2018}. This result is consistent with the arguments made in \citet[][]{Acuner2018} that these bursts could be due to synchrotron emission. However, dissipation below the photosphere could also produce broad spectra. Therefore, physical model comparison should be done in order to assess the underlying physics for these two clusters.

\begin{table*}
	
    \centering
	\caption{Distributions of $\alpha$ for  synthetic photospheric spectra (coasting phase and no dissipation). Simulations were made with $E_{\rm pk}$-values sampled from the distributions in the GBM catalogue, and in clusters 1 to 5 defined in \citet{Acuner2018}. The parameters are the medians and inter-quantile ranges (IQR). }
	\label{tab:2}

    \begin{tabular}{llllll}
     Sample   & ~        &~    & $\alpha_{median}$     &           IQR       \\   
     \hline
     Full GBM  & ~     &~                  &  -0.16 & 0.15                                            \\ \hline
    Cluster 1 & ~     &~                  &   -0.13 & 0.044                                             \\ \hline
    Cluster 2 &~    &~                     &  -0.13 & 0.09                                              \\ \hline
    Cluster 3 &~    &~                     &  -0.18 & 0.068                                              \\ \hline
    Cluster 4 &~    &~                     & -0.2 & 0.113                                              \\ \hline
    Cluster 5 & ~    &~                & -0.08 &  0.043                                                \\ 
		\hline
	\end{tabular}
\end{table*}

\subsection{Comparison to the maximal, time-resolved $\alpha$-value in each burst}
\label{sec:amax}
In the previous sections we have compared the predicted range of $\alpha$-values from NDP to the observed distributions of $\alpha$ from the GBM catalogue.
As mentioned above, these spectra are from the timebin at the flux peak for each burst, integrating over $\sim 1$ s \citep{Gruber2014}. Hence, this selection yields spectra that are not necessarily time-resolved enough, since the light curve variability is not considered. The five clusters discussed in \S~\ref{sec:all} all have different variability properties \citep[Table 2 in][]{Acuner2018}. In particular, we note that the median variability time for cluster 3 and 4 are around 1 second while for the other clusters it is $< 0.5$ seconds. Therefore, the analysed timebins in clusters 3 and 4 might be time-resolved enough to exhibit the instantaneous emission spectrum.

On the other hand, time-resolved spectral catalogues have been presented, but they are for significantly smaller samples \citep[e.g, ][]{Kaneko2006, Yu2016, Yu2019}. Furthermore, the parameter distribution in these catalogues contain values from all the time-resolved spectra. As a consequence, these distributions contain varying number of spectra from each individual burst.  This will give rise to a bias towards long and strong bursts, which have many time bins and therefore will dominate the $\alpha$-distribution. In addition, since spectral evolution is expected during individual bursts, the distributions also reflect the evolution in $\alpha$, even though there is no spectral evolution within timebins. The latter point is important since there are bursts which initially are unquestionably photospheric, but later evolve into a broader spectrum that can be fitted by non-thermal emission models \citep[see, e.g.][]{Ryde2011, Guiriec2015b, Yu2019, ZhangBB2018, Ryde2019}.  Indeed, since emission processes  typically are easiest distinguishable  by the hardest $\alpha$-value that they allow, and if one assumes a single emission mechanism throughout a bursts, then  $\alpha_{\rm max}$ is the most indicative parameter for the burst emission.
In order to constrain the emission process during bursts, it should, therefore,  be  more informative to study the distribution of the maximal value of $\alpha$ in every individual burst instead.

\begin{figure}
 \includegraphics[width=\columnwidth]{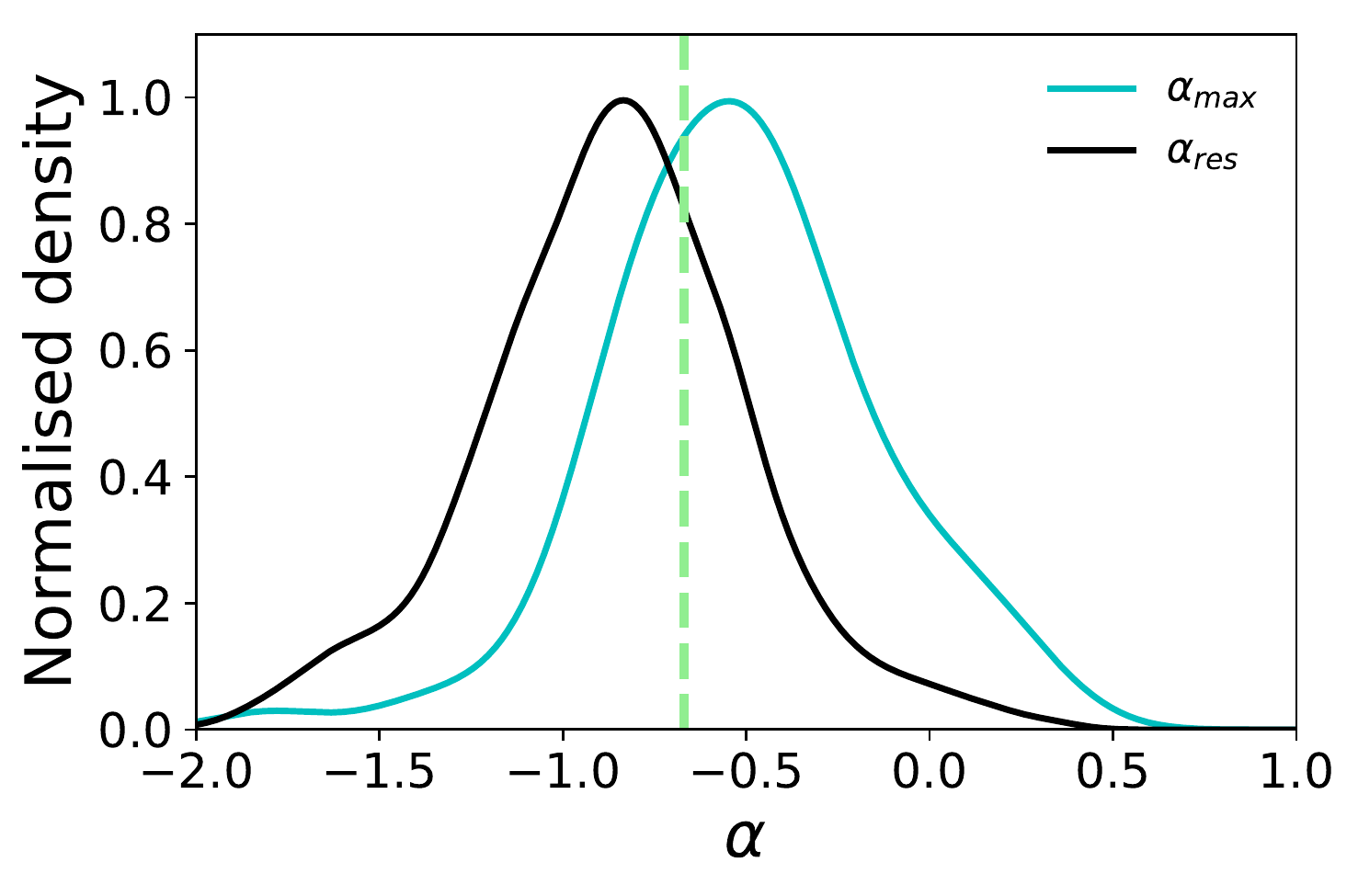}
 \caption{Normalised density distributions of $\alpha$-values from the 81 bursts analysed in the \citet{Yu2016} GBM catalogue, as KDE (Kernel Density Estimation) curves. The blue distribution is for the maximal $\alpha$-values in each of the 81 bursts ($\alpha_{\rm max}$). In comparison, the black distribution shows the KDE curve of all the 1754 time-resolved $\alpha$-values from the same 81 bursts ($\alpha_{res}$). The green line shows the $\alpha=-2/3$.}
\label{fig:yualphamax}
\end{figure}

To illustrate this, we plot in  Figure \ref{fig:yualphamax} the distribution for  $\alpha_{\rm max}$ for the 81 bursts in the GBM time-resolved catalogue \citep{Yu2016}. The time resolved distribution has an $\alpha$ median of -0.84, whereas the median for $\alpha_{\rm max}$ is -0.53. As can be seen from the figure, the distribution of $\alpha$ is significantly shifted to harder values. In particular, a majority of the spectra are now beyond the line-of-death for synchrotron emission, $\alpha = -2/3$ \citep[green dashed line; ][]{Preece1998}). The fraction changes from 29\% to 65\%. 
The $\alpha_{\rm max}$-distribution in Figure \ref{fig:yualphamax} is similar to the corresponding distribution for the sample of 38 single pulses in \citet{Yu2019}; see their Fig. 3, which reaffirms the result. The significant change in the $\alpha$-distribution thus calls for a separate investigation of the $\alpha_{\rm max}$-timebins.

\begin{figure}
 \includegraphics[width=\columnwidth]{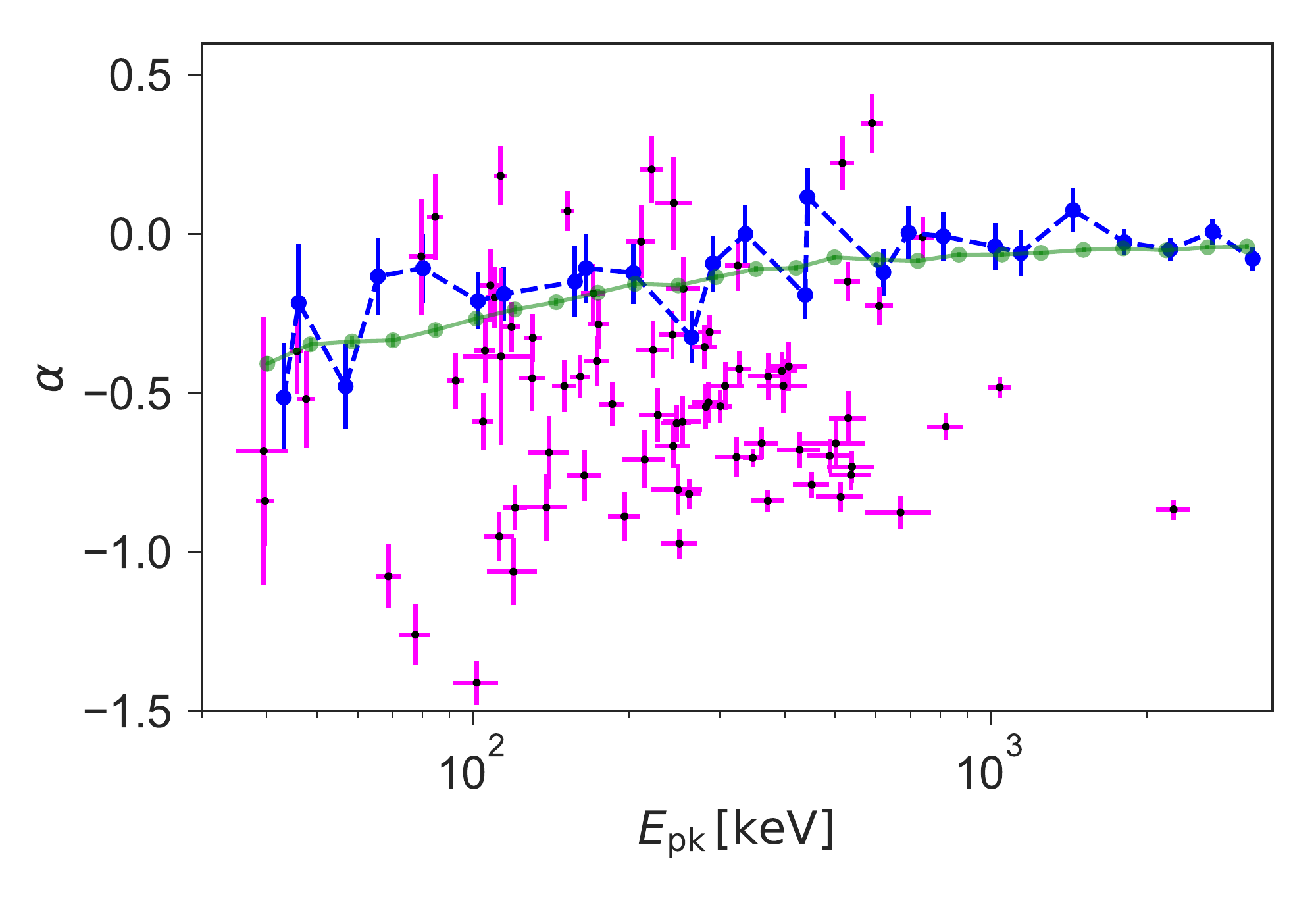}
 \includegraphics[width=\columnwidth]{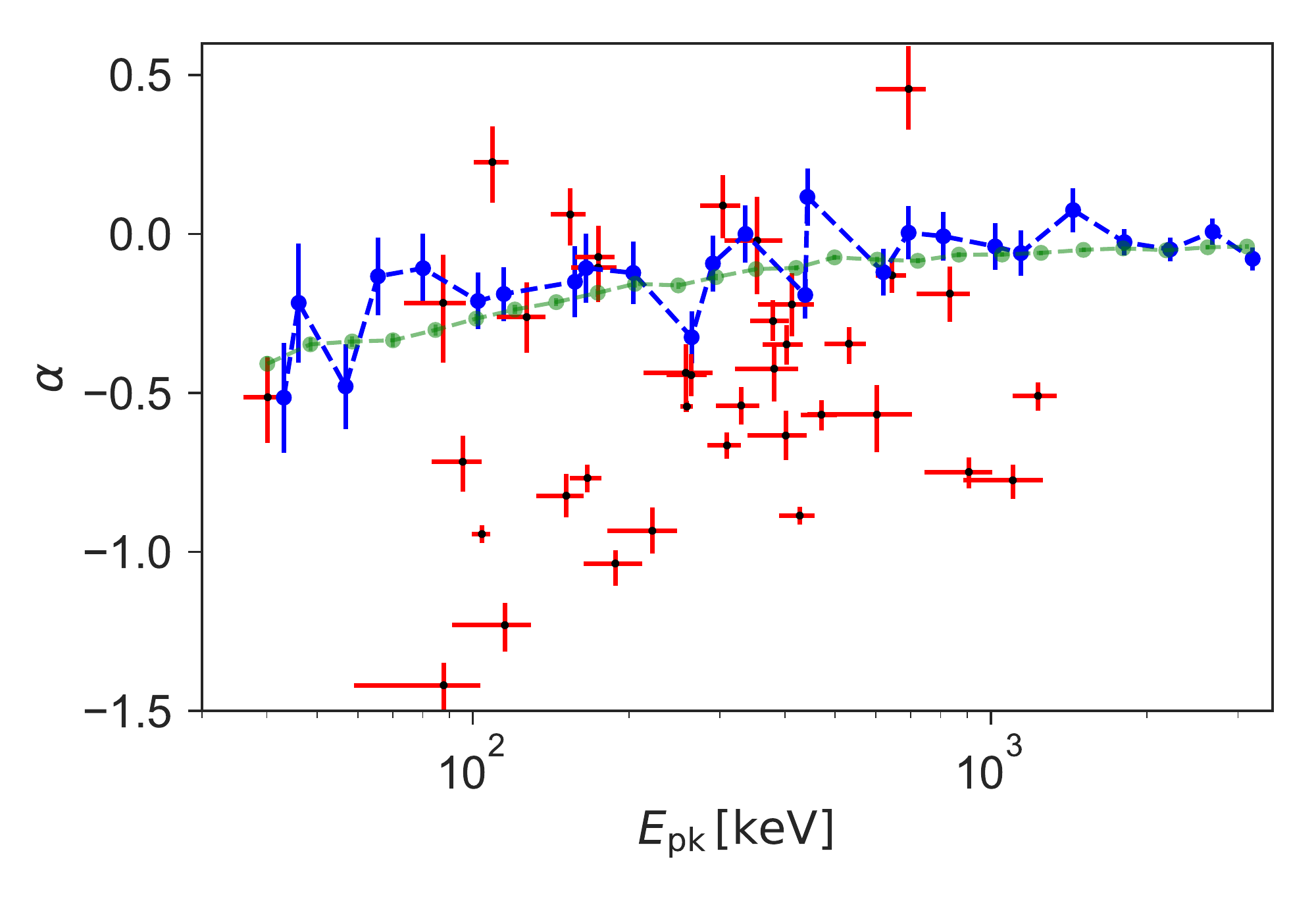}
 \caption{Distribution of observed ($\alpha_{\rm max}$, $E_{\rm pk}$) over-plotted the expectations from synthetic non-dissipative photosphere spectra (blue and green lines; see Fig. \ref{fig:data}). Upper panel:  Magenta data points are from the time-resolved GBM catalogue \citep{Yu2016}. Around 26\% of the observed points (21/81) are consistent with the synthetic burst spectra, since they lie above or within 1$\sigma$ below the green line. 
Lower panel: Red data points are for the GRBs in the \citet{Yu2018} pulse catalogue. Here again around 29\% of the observed bursts (11/38) 
are consistent.
  }
 \label{fig:alphamaxGBM}
\end{figure}

In the upper panel in Figure \ref{fig:alphamaxGBM}, we plot the  distribution of $\alpha_{\rm max}$ and their corresponding $E_{\rm pk}$-values, from the 81 bursts in the GBM time-resolved catalogue \citep{Yu2016}. The expected values for non-dissipative photospheres is shown by the fits to synthetic bursts (blue and green lines from Figure \ref{fig:data}.)
We find that around 26\% of the observed points (21/81) are consistent with the synthetic burst spectra, since they cover the same region in the $\alpha$--$E_{\rm pk}$--plane: They lie above or within $1\sigma$ below the green line. In the lower panel in Figure \ref{fig:alphamaxGBM}, the corresponding relation is plotted for the 38 pulses in \citet{Yu2019}. Here, we find that around 29\% of the observed points (11/38) are consistent. 

While some overlap exists between the two samples (\citet{Yu2016} and \citet{Yu2019}), we note that they are different in how the bursts were selected. Bursts in  \citet{Yu2019} are selected to be single pulses, while \citet{Yu2016} also allowed  multi-pulse, and complex bursts. Furthermore, \citet{Yu2019} uses  the Bayesian Blocks method \citep{Scargle2013}  
for creating the timebins, to which they later apply an significance threshold \citep{Vianello2018},  
whereas \citet{Yu2016} only utilizes a signal-to-noise ratio threshold to determine the timebins. 

Therefore, we argue that the clearest signature of the emission mechanism will be provided by the analysis of the \citet{Yu2019}-sample, depicted in Figure \ref{fig:alphamaxGBM}. This analysis indicates that around 29\% of the bursts have at least one timebin with a spectrum that is consistent with a non-dissipative photosphere. 

\section{Discussion} 

We have studied the observed properties of photospheric emission in the specific case where the photon distributions are unaltered by energy dissipation in the flow, the so called non-dissipative photospheres. If such spectra are observed by a $\gamma$-ray instrument with a limited energy-band and are fitted by the typically employed empirical functions, then these spectra will yield a distribution of $\alpha$-values, lying in the range $-0.4 < \alpha < 0.0$. This is in contrast to the generally expected value of $\alpha = 0.4$ for non-dissipative  photospheres.

\subsection{Explaining the observed $\alpha$-distributions}
\label{sec:alpha_distribution}

A consequence of this is that a non-negligible fraction of  the $\alpha$-distribution from bursts observed by, for instance, the GBM, are compatible with  being from non-dissipative photospheres.  Earlier, only  a handful of bursts have been claimed to be from NDP 
\citep{Ryde2004, Ghirlanda2013, Larsson2015}. \citet{Ryde2017} analysed  two  of these bursts, which exhibit extremely narrow spectra, namely GRB100507 and GRB101219. The spectral evolution were indeed consistent with a photosphere transitioning from the acceleration to the coasting phase of the flow, a possibility also discussed in \citet{Chhotray2018}. Here, our results indicate that many more bursts could be interpreted as stemming from such photospheres, just based on their $\alpha$-values (Fig. \ref{fig:all}). 

This is remarkable, since non-dissipative photospheric emission is only expected in very special circumstances, such as very smooth flows. The reason is that during the coasting phase the jet kinetic energy dominates the radiation energy. Therefore, dissipating only a small fraction of the kinetic energy can easily yield an energy density that is comparable to the energy density of the photon field prior to the dissipation. In such a situation the spectrum from the photosphere is expected to be modified from the shape in Eq. \ref{eq:1} \citep[e.g., ][]{Rees&Meszaros2005, Peer2006,Giannios2006, Vurm2011, ahlgren2015confronting}. 
Above the spectral peak, the spectrum can become harder due to Comptonization, while below the peak, low-energy, synchrotron photons, that are produced at the dissipation site, can make the spectrum softer. The change in the spectrum below the peak will cause smaller $\alpha$-values. The absence of such broadening thus places strong constraints on the amount of dissipation and photon production that is allowed. 

It is thus a natural expectation that a part of the kinetic energy of the flow below the photosphere is allowed to dissipate in many bursts. 
A consequence of this is that an even larger fraction of the $\alpha$-value distribution could be consistent with emission from the photosphere. 
For instance, bursts that have  $\alpha_{\rm max} < -0.4$ (limiting value for NDP) will then be compatible with a photospheric scenario as well. 

This could also explain the distributions shown in Appendix C, in particular the
limited overlaps between the observed and simulated $\alpha$-values  (see Table \ref{tab:1}).  The fraction of bursts overlapping are from bursts not affected by dissipation, while the others are affected. The characteristics of the observed photospheric emission,  would then depend on the individual properties of the flow yielding a varying amount of dissipation.

On the other hand, the complementary fraction of spectra in the $\alpha$-distribution  ($\alpha < -0.4$) could be due to a wholly different emission process, such as synchrotron emission \citep[see, e.g., ][]{Beniamini&Piran2013, Acuner2018}. 
In such a case one could imagine that during an initial phase of a burst, the emission from the photosphere dominates upon which synchrotron emission supervenes from a different part of the flow, either from internal or external shocks \cite[e.g., ][]{Ghirlanda2004, ZhangBB2018, Li2018}. However, the evolution of spectral parameters are typically smooth, which questions a two-zone emission scenario in which the two zones supplant in dominance of the emission \citep[e.g., ][]{Ryde2019}.

\subsection{High-energy spectral shape and bursts in cluster 3}
\label{sec:HE}


 Low-energy indices are widely used to interpret radiation mechanisms from GRBs because of their distinguishability between various models. In this study, we concentrated on the $\alpha$-values which determine the seed radiative process that defines the lower energy tail of the spectra. 
 
 On the other hand, significant information also lies in the spectral shape above the peak energy, which therefore should not be neglected whilst inferring the emission physics. For instance, 
 a high-energy power-laws can be the result of  thermal Comptonization \citep{RybickiLightman}, inverse Compton scattering \citep{SternPoutanen2004}, or thermal synchrotron emission \citep[e.g, ][]{Petrosian1981}.
 
 However, spectral analysis in large samples of GRBs suggest that the best empirical model is the cut-off powerlaw function and that the data do not require a high-energy power-law \citep{Goldstein2012, Gruber2014, Yu2016}. In particular, \citet{Yu2019} showed that if the analysed timebins are chosen to avoid spectral evolution, and therefore can be considered to exhibit the instantaneous emission spectrum, then the Band function $\beta$ becomes softer (more negative) and thus more compatible with an exponential cut-off. In such cases, the $\alpha$-value of the spectra is the decisive parameter and the conclusions on the fraction of bursts consistent with NDP spectra based on the $\alpha$-distribution are valid.
 
On the other hand, a particular feature of the spectra in one of the clusters in \citet{Acuner2018} (cluster 3) are very hard values of the high-energy, powerlaw index $\beta \sim -2$. The presence of such a powerlaw indicates that energy dissipation has occurred and that the Compton $y$-parameter is $y \sim 1$. The high-energy, power-law emission could either be due to optically-thin Compton scattering, which would give rise to a wavy high-energy spectrum (as indicated in \citet{Acuner2018}) or optically thick Comptonisation, producing a pure power-law. In either case, such a feature also indicates  that the flow has dissipated an amount of energy which is approximately of the same amount as the thermal photon energy.

Interestingly, we found above (\S \ref{sec:all}) that, comparing clusters 1, 3, and 5  \citep[photosphere clusters in][]{Acuner2018}, cluster 3 differs in that it has  the largest overlap between the measured and the expected NDP $\alpha$-values (see Table \ref{tab:1} and Appendix C). However, even though a large fraction of the bursts in cluster 3 have  $\alpha$-values that correspond to the expectation of NDP spectra, they are obviously not all non-dissipative, due to the presence of a high energy tail in the spectra. 

This could therefore, in part, explain the fact that cluster 3 has a larger overlap
between the measured and the expected NDP $\alpha$-values, compared to  clusters 1 and 5. The fraction of spectra affected by dissipation might be similar between the three clusters, however, the effects of the dissipation could be different. In clusters  1 and  5, the dissipation mainly affects the low-energy part ($\alpha$) of the spectra: The dissipation produces low-energy (synchrotron) photons which are energised and populate the spectrum below the peak. This results in that a majority of all spectra are not compatible with the NDP.
In contrast, the effect of the dissipation in the cluster 3-bursts could be different. For instance, for some of the burst, the low-energy photon production could be suppressed,
 leaving the $\alpha$-slope is unchanged. At the same time, the energy dissipation occurring in these bursts would mainly cause changes in the high-energy part of the spectrum. Thus, the actual fraction of bursts in cluster 3 without dissipation, i.e., truely non-dissipative, should be lower than what is given in Table \ref{tab:1}.


 
 
As a conclusion, only focusing on $\alpha$-values, therefore,  does not necessarily reveal the actual presence of dissipation. An investigation of the low-energy index should therefore be followed by an investigation of  the whole spectrum, in order to address the effects of Comptonisation.
 This requires different physical models to be fit to data and to be compared through a statistically reasonable methods. 

\subsection{Interpretation of $\alpha$ and extrapolation to the X-ray spectrum}

Figure \ref{fig:moda} shows that the fitting process of minimizing the fit statistics of the cut-off powerlaw function, results in a fit that only correctly determines the spectral curvature above $\sim 35$ keV. However, the fit is not able to correctly  reproduce the actual powerlaw slope at low energies. The  $\alpha$ parameter attains a value which optimizes the match to the  actual spectral curvature around the peak and therefore it does not necessarily fit the spectral shape below $\sim 35$  keV.

This can in part be understood by the significance of the counts per channel for the synthetic photosphere spectra.  The significance of the data will be the largest at a few 100 keV. This is partly due to the decrease in effective area towards lower energies and partly due to the subpeak powerlaw slope of the photon spectrum (Eq. \ref{eq:1}) having a slope steeper than $N_{\rm E} \propto E^{0}$, that is, an increasing photon number with energy. For typical spectra the significance can vary by a factor of four. Therefore, more weight in the fit is given to the count data around a few 100 keV, even baring in mind that dispersion of incoming photons occur in the detector to produce the count spectrum \citep{Meegan2009_GBM}. 
As a consequence the data far below the peak will have less influence on the over all fit. Hence, the fitted $\alpha$-value does not necessarily represent the asymptotic slope of the incoming spectrum.


In such a case, 
 $\alpha$ should be interpreted as a parameter that is used to fit the spectral curvature around the peak, and not the asymptotic powerlaw slope of the incoming spectrum, which is typically done. 
Hence, the determined value of $\alpha$ should, for instance, not be used to 
extrapolate the GBM data to lower energies, such as, the X-ray or optical bands \citep{ Ghirlanda2007, Sakamoto2009}. The residuals from the spectrum in Figure \ref{fig:moda} can become large. For instance, at 1 keV,  their ratio is 0.2. This means that if (i) the true spectrum is a NDP, and (ii)  if an extrapolation were to be made into the X-ray energy range and below, based on the $\alpha$-value alone, the spectral flux will be overestimated by nearly an order of magnitude. Note that the parameter $\alpha$ is not necessarily badly determined statistically. It is the interpretation of it that might be wrong, simply due to the fitted model not being the accurate model, which, in this case, is assumed to be the NDP.

\subsection{Determination of $E_{pk}$}

\begin{figure}
 \includegraphics[width = \columnwidth]{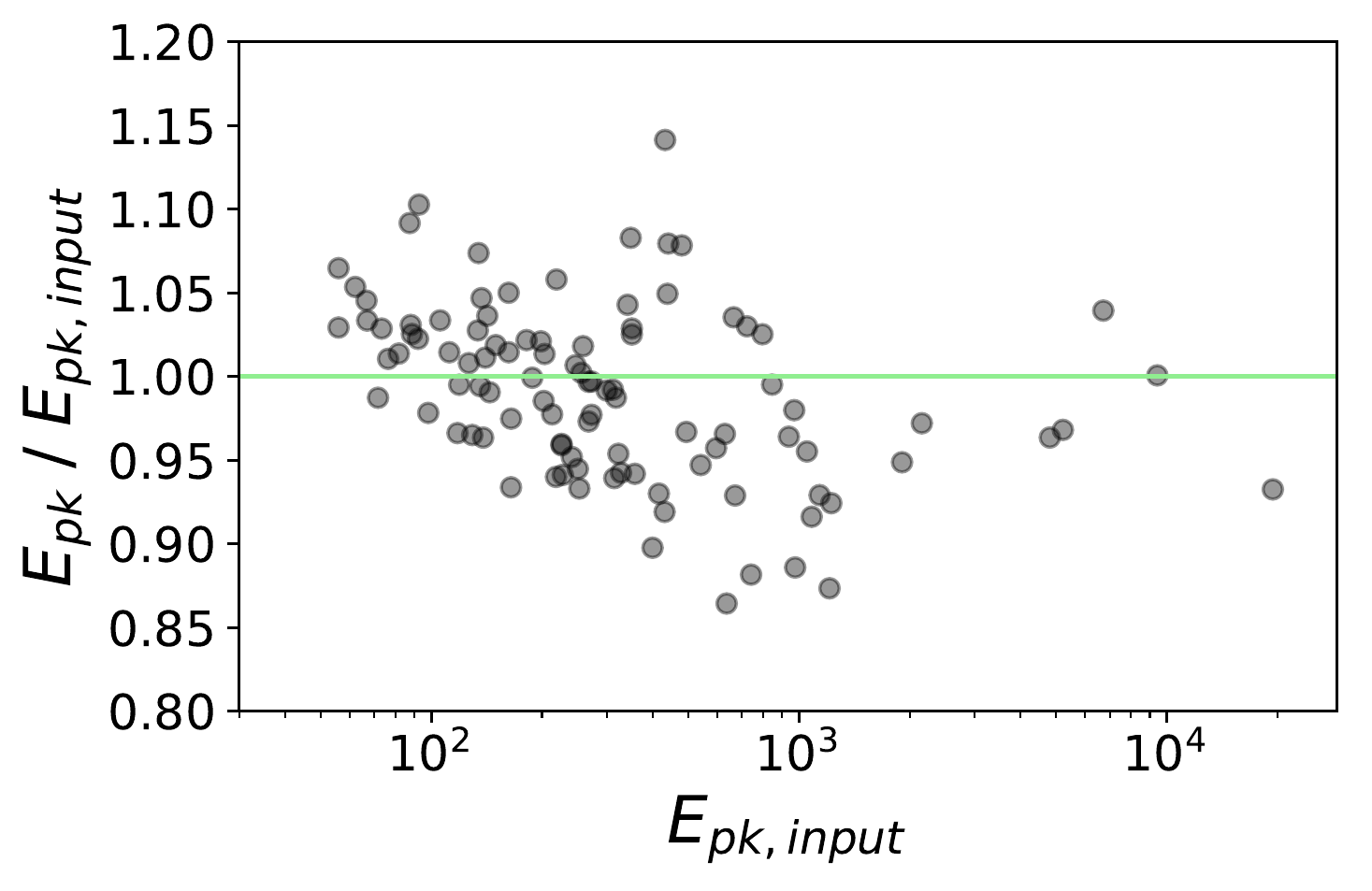}
 \caption{Ratio between the fitted value ($E_{\rm pk}$) and the correct value ($E_{\rm pk,\, input}$) versus the correct value ($E_{\rm pk,\, input}$). 
 This shows that the energy of the spectral break is well determined.
 }
 \label{fig:epkdet}
\end{figure}

In the catalogue simulations made above, we have sampled the $E_{\rm pk}$-values that we use for the NDP spectra, from the $E_{\rm pk}$-distribution found from the cut-off powerlaw fits. This assumes that the  $E_{\rm pk}$ is well determined by the cut-off powerlaw.
On the other hand,  the fitted $\alpha$-values can be quite off from the expected value. This is  shown in  Figure \ref{fig:all}, where the determined $\alpha$-values are shown by the blue line, which is significantly off-set from the asymptotic power law slope ($\alpha = 0.4$) of the NDP spectra, which  is shown by the blue, dashed line. 
However, Figure \ref{fig:epkdet} shows how well the $E_{\rm pk}$ is determined by 
showing the ratio of $E_{\rm pk}$ and $E_{\rm pk, input}$ versus $E_{\rm pk, input}$. The ratio is consistently close to 1.0, and there is no strong trend. This, therefore, shows that there is no disadvantage of sampling from the fitted $\Epk$ from using the  cut-off powerlaw function. 

\section{Conclusion}

We have shown that GRB spectra that have $\alpha > -0.4$ have low-energy slopes that are compatible with a photospheric emission from a non-dissipative outflow. This corresponds to a non-negligible fraction of all GRB spectra (Fig. \ref{fig:all}). The reason one should not expect a hard $\alpha$-value from photospheric emission, such as $\alpha =1$ (Rayleigh-Jeans slope) or $\alpha =0.4$ (including geometrical effects)  is that the empirical models used in spectral analysis (cut-off powerlaw and Band) do not match the curvature of the incoming spectrum. A consequence of this is that the determined value of $\alpha$ will be consistently softer than the asymptotic slope of the theoretical spectrum ($\alpha \sim 0.4$). The fitted value is instead closer to $\alpha \sim -0.1$.

We find that more than a quarter of all bursts have at least one time-resolved spectrum, whose $\alpha$-values are consistent with a non-dissipative outflow photosphere. This is the most restrictive and simplest photospheric scenario. However, since it is natural to expect a degree of dissipation of kinetic energy in the flow  below the photosphere, the fraction of spectra consistent with emission from the photosphere should be even higher.


The conclusion is that 
(i) a non-negligible fraction of the observed $\alpha$ distribution can be explained by a non-dissipative photospheric model. (ii) It is not advisable to extrapolate the determined value of $\alpha$ to lower energy bands, such as the X-ray and the optical bands. 
(iii) The interpretation of $\alpha$-values to assess emission models can be misleading and must be done with caution.










\section*{Acknowledgements}

We thank Dr. Lundman for enlightening discussions.
We acknowledge support from the Swedish National Space Agency and the Swedish Research Council (Vetenskapsr{\aa}det). 
FR is supported by the  G\"oran Gustafsson Foundation for Research in Natural Sciences and Medicine. 




\bibliographystyle{mnras}
\bibliography{ref2017} 

\begin{thebibliography}{}
\makeatletter
\relax
\def\mn@urlcharsother{\let\do\@makeother \do\$\do\&\do\#\do\^\do\_\do\%\do\~}
\def\mn@doi{\begingroup\mn@urlcharsother \@ifnextchar [ {\mn@doi@}
  {\mn@doi@[]}}
\def\mn@doi@[#1]#2{\def\@tempa{#1}\ifx\@tempa\@empty \href
  {http://dx.doi.org/#2} {doi:#2}\else \href {http://dx.doi.org/#2} {#1}\fi
  \endgroup}
\def\mn@eprint#1#2{\mn@eprint@#1:#2::\@nil}
\def\mn@eprint@arXiv#1{\href {http://arxiv.org/abs/#1} {{\tt arXiv:#1}}}
\def\mn@eprint@dblp#1{\href {http://dblp.uni-trier.de/rec/bibtex/#1.xml}
  {dblp:#1}}
\def\mn@eprint@#1:#2:#3:#4\@nil{\def\@tempa {#1}\def\@tempb {#2}\def\@tempc
  {#3}\ifx \@tempc \@empty \let \@tempc \@tempb \let \@tempb \@tempa \fi \ifx
  \@tempb \@empty \def\@tempb {arXiv}\fi \@ifundefined
  {mn@eprint@\@tempb}{\@tempb:\@tempc}{\expandafter \expandafter \csname
  mn@eprint@\@tempb\endcsname \expandafter{\@tempc}}}

\bibitem[\protect\citeauthoryear{{Abramowicz}, {Novikov}  \&
  {Paczynski}}{{Abramowicz} et~al.}{1991}]{Abramowicz1991}
{Abramowicz} M.~A.,  {Novikov} I.~D.,   {Paczynski} B.,  1991, \mn@doi [ApJ]
  {10.1086/169748}, \href {http://adsabs.harvard.edu/abs/1991ApJ...369..175A}
  {369, 175}

\bibitem[\protect\citeauthoryear{{Acuner} \& {Ryde}}{{Acuner} \&
  {Ryde}}{2018}]{Acuner2018}
{Acuner} Z.,  {Ryde} F.,  2018, \mn@doi [\mnras] {10.1093/mnras/stx3106}, \href
  {http://adsabs.harvard.edu/abs/2018MNRAS.475.1708A} {475, 1708}

\bibitem[\protect\citeauthoryear{Ahlgren, Larsson, Nymark, Ryde  \&
  Pe'er}{Ahlgren et~al.}{2015}]{ahlgren2015confronting}
Ahlgren B.,  Larsson J.,  Nymark T.,  Ryde F.,   Pe'er A.,  2015, Monthly
  Notices of the Royal Astronomical Society: Letters, 454, L31

\bibitem[\protect\citeauthoryear{{Ahlgren}, {Larsson}, {Ahlberg}, {Lundman},
  {Ryde}  \& {Pe'er}}{{Ahlgren} et~al.}{2019}]{Ahlgren2019}
{Ahlgren} B.,  {Larsson} J.,  {Ahlberg} E.,  {Lundman} C.,  {Ryde} F.,
  {Pe'er} A.,  2019, arXiv e-prints, \href
  {http://adsabs.harvard.edu/abs/2019arXiv190106844A} {}

\bibitem[\protect\citeauthoryear{{Arnaud}}{{Arnaud}}{1996}]{Arnaud1996}
{Arnaud} K.~A.,  1996, in {Jacoby} G.~H.,  {Barnes} J.,  eds,  Astronomical
  Society of the Pacific Conference Series Vol. 101, Astronomical Data Analysis
  Software and Systems V. p.~17

\bibitem[\protect\citeauthoryear{{Axelsson}, {Baldini}  \& et al.}{{Axelsson}
  et~al.}{2012}]{Axelsson2012}
{Axelsson} M.,  {Baldini} L.,   et al. 2012, \mn@doi [ApJ]
  {10.1088/2041-8205/757/2/L31}, \href
  {http://adsabs.harvard.edu/abs/2012ApJ...757L..31A} {757, L31}

\bibitem[\protect\citeauthoryear{{Band}, {Matteson}  \& et al.}{{Band}
  et~al.}{1993}]{Band1993}
{Band} D.,  {Matteson} J.,   et al. 1993, \mn@doi [ApJ] {10.1086/172995}, \href
  {http://adsabs.harvard.edu/abs/1993ApJ...413..281B} {413, 281}

\bibitem[\protect\citeauthoryear{{Beloborodov}}{{Beloborodov}}{2010}]{Beloborodov2010}
{Beloborodov} A.~M.,  2010, \mn@doi [MNRAS] {10.1111/j.1365-2966.2010.16770.x},
  \href {http://adsabs.harvard.edu/abs/2010MNRAS.407.1033B} {407, 1033}

\bibitem[\protect\citeauthoryear{{Beloborodov}}{{Beloborodov}}{2011}]{Beloborodov2011}
{Beloborodov} A.~M.,  2011, \mn@doi [ApJ] {10.1088/0004-637X/737/2/68}, \href
  {http://adsabs.harvard.edu/abs/2011ApJ...737...68B} {737, 68}

\bibitem[\protect\citeauthoryear{{Beniamini} \& {Piran}}{{Beniamini} \&
  {Piran}}{2013}]{Beniamini&Piran2013}
{Beniamini} P.,  {Piran} T.,  2013, \mn@doi [ApJ] {10.1088/0004-637X/769/1/69},
  \href {http://adsabs.harvard.edu/abs/2013ApJ...769...69B} {769, 69}

\bibitem[\protect\citeauthoryear{{Burgess}, {Ryde}  \& {Yu}}{{Burgess}
  et~al.}{2015}]{Burgess2015_alpha}
{Burgess} J.~M.,  {Ryde} F.,   {Yu} H.-F.,  2015, \mn@doi [\mnras]
  {10.1093/mnras/stv775}, \href
  {http://adsabs.harvard.edu/abs/2015MNRAS.451.1511B} {451, 1511}

\bibitem[\protect\citeauthoryear{{Burgess}, {B{\'e}gu{\'e}}, {Bacelj},
  {Giannios}, {Berlato}  \& {Greiner}}{{Burgess} et~al.}{2018}]{Burgess2018}
{Burgess} J.~M.,  {B{\'e}gu{\'e}} D.,  {Bacelj} A.,  {Giannios} D.,  {Berlato}
  F.,   {Greiner} J.,  2018, arXiv e-prints, \href
  {http://adsabs.harvard.edu/abs/2018arXiv181006965B} {}

\bibitem[\protect\citeauthoryear{{Chhotray} \& {Lazzati}}{{Chhotray} \&
  {Lazzati}}{2018}]{Chhotray2018}
{Chhotray} A.,  {Lazzati} D.,  2018, \mn@doi [\mnras] {10.1093/mnras/sty286},
  \href {http://adsabs.harvard.edu/abs/2018MNRAS.476.2352C} {476, 2352}

\bibitem[\protect\citeauthoryear{De~Colle, Lu, Kumar, Ramirez-Ruiz  \&
  Smoot}{De~Colle et~al.}{2017}]{DeColle2017}
De~Colle F.,  Lu W.,  Kumar P.,  Ramirez-Ruiz E.,   Smoot G.,  2017, arXiv.org,
  p. arXiv:1701.05198

\bibitem[\protect\citeauthoryear{{Dermer} \& {B{\"o}ttcher}}{{Dermer} \&
  {B{\"o}ttcher}}{2000}]{DermerBottcher2000}
{Dermer} C.~D.,  {B{\"o}ttcher} M.,  2000, \mn@doi [\apjl] {10.1086/312669},
  \href {http://adsabs.harvard.edu/abs/2000ApJ...534L.155D} {534, L155}

\bibitem[\protect\citeauthoryear{Gelman, Carlin, Stern, Dunson, Vehtari  \&
  Rubin}{Gelman et~al.}{2014a}]{bda3}
Gelman A.,  Carlin J.~B.,  Stern H.,  Dunson D.~B.,  Vehtari A.,   Rubin D.~B.,
   2014a, Bayesian data analysis.
CRC Press

\bibitem[\protect\citeauthoryear{Gelman, Hwang  \& Vehtari}{Gelman
  et~al.}{2014b}]{Gelman2014}
Gelman A.,  Hwang J.,   Vehtari A.,  2014b, \mn@doi [Statistics and Computing]
  {10.1007/s11222-013-9416-2}, 24, 997

\bibitem[\protect\citeauthoryear{{Ghirlanda}, {Ghisellini}  \&
  {Celotti}}{{Ghirlanda} et~al.}{2004}]{Ghirlanda2004}
{Ghirlanda} G.,  {Ghisellini} G.,   {Celotti} A.,  2004, \mn@doi [\aap]
  {10.1051/0004-6361:20048008}, \href
  {http://adsabs.harvard.edu/abs/2004A%26A...422L..55G} {422, L55}

\bibitem[\protect\citeauthoryear{{Ghirlanda}, {Bosnjak}  \& et al.}{{Ghirlanda}
  et~al.}{2007}]{Ghirlanda2007}
{Ghirlanda} G.,  {Bosnjak} Z.,   et al. 2007, \mn@doi [MNRAS]
  {10.1111/j.1365-2966.2007.11890.x}, \href
  {http://adsabs.harvard.edu/abs/2007MNRAS.379...73G} {379, 73}

\bibitem[\protect\citeauthoryear{{Ghirlanda}, {Ghisellini}  \& et
  al.}{{Ghirlanda} et~al.}{2013}]{Ghirlanda2013}
{Ghirlanda} G.,  {Ghisellini} G.,   et al. 2013, \mn@doi [MNRAS]
  {10.1093/mnras/sts128}, \href
  {http://adsabs.harvard.edu/abs/2013MNRAS.428.1410G} {428, 1410}

\bibitem[\protect\citeauthoryear{{Giannios}}{{Giannios}}{2006}]{Giannios2006}
{Giannios} D.,  2006, \mn@doi [A{$\&$}A] {10.1051/0004-6361:20065000}, \href
  {http://adsabs.harvard.edu/abs/2006A%26A...457..763G} {457, 763}

\bibitem[\protect\citeauthoryear{{Giannios} \& {Uzdensky}}{{Giannios} \&
  {Uzdensky}}{2019}]{Giannios2019}
{Giannios} D.,  {Uzdensky} D.~A.,  2019, \mn@doi [\mnras]
  {10.1093/mnras/stz082}, \href
  {http://adsabs.harvard.edu/abs/2019MNRAS.484.1378G} {484, 1378}

\bibitem[\protect\citeauthoryear{{Goldstein} et~al.,}{{Goldstein}
  et~al.}{2012}]{Goldstein2012}
{Goldstein} A.,  et~al., 2012, \mn@doi [\apjs] {10.1088/0067-0049/199/1/19},
  \href {http://adsabs.harvard.edu/abs/2012ApJS..199...19G} {199, 19}

\bibitem[\protect\citeauthoryear{{Golkhou} \& {Butler}}{{Golkhou} \&
  {Butler}}{2014}]{Golkhou&Butler2014}
{Golkhou} V.~Z.,  {Butler} N.~R.,  2014, \mn@doi [ApJ]
  {10.1088/0004-637X/787/1/90}, \href
  {http://adsabs.harvard.edu/abs/2014ApJ...787...90G} {787, 90}

\bibitem[\protect\citeauthoryear{{Golkhou}, {Butler}  \&
  {Littlejohns}}{{Golkhou} et~al.}{2015}]{Golkhou&Butler2015}
{Golkhou} V.~Z.,  {Butler} N.~R.,   {Littlejohns} O.~M.,  2015, \mn@doi [\apj]
  {10.1088/0004-637X/811/2/93}, \href
  {http://adsabs.harvard.edu/abs/2015ApJ...811...93G} {811, 93}

\bibitem[\protect\citeauthoryear{{Goodman}}{{Goodman}}{1986}]{Goodman1986}
{Goodman} J.,  1986, \mn@doi [ApJL] {10.1086/184741}, \href
  {http://adsabs.harvard.edu/abs/1986ApJ...308L..47G} {308, L47}

\bibitem[\protect\citeauthoryear{{Goodman} \& {Weare}}{{Goodman} \&
  {Weare}}{2010}]{emcee}
{Goodman} J.,  {Weare} J.,  2010, \mn@doi [Communications in Applied
  Mathematics and Computational Science, Vol.~5, No.~1, p.~65-80, 2010]
  {10.2140/camcos.2010.5.65}, \href
  {http://adsabs.harvard.edu/abs/2010CAMCS...5...65G} {5, 65}

\bibitem[\protect\citeauthoryear{{Gruber} et~al.,}{{Gruber}
  et~al.}{2014}]{Gruber2014}
{Gruber} D.,  et~al., 2014, \mn@doi [\apjs] {10.1088/0067-0049/211/1/12}, \href
  {http://adsabs.harvard.edu/abs/2014ApJS..211...12G} {211, 12}

\bibitem[\protect\citeauthoryear{{Guiriec} et~al.,}{{Guiriec}
  et~al.}{2013}]{Guiriec2013}
{Guiriec} S.,  et~al., 2013, \mn@doi [ApJ] {10.1088/0004-637X/770/1/32}, \href
  {http://adsabs.harvard.edu/abs/2013ApJ...770...32G} {770, 32}

\bibitem[\protect\citeauthoryear{{Guiriec}, {Mochkovitch}, {Piran}, {Daigne},
  {Kouveliotou}, {Racusin}, {Gehrels}  \& {McEnery}}{{Guiriec}
  et~al.}{2015}]{Guiriec2015b}
{Guiriec} S.,  {Mochkovitch} R.,  {Piran} T.,  {Daigne} F.,  {Kouveliotou} C.,
  {Racusin} J.,  {Gehrels} N.,   {McEnery} J.,  2015, \mn@doi [\apj]
  {10.1088/0004-637X/814/1/10}, \href
  {http://adsabs.harvard.edu/abs/2015ApJ...814...10G} {814, 10}

\bibitem[\protect\citeauthoryear{{Ito} et~al.,}{{Ito} et~al.}{2013}]{Ito2013}
{Ito} H.,  et~al., 2013, \mn@doi [\apj] {10.1088/0004-637X/777/1/62}, \href
  {http://adsabs.harvard.edu/abs/2013ApJ...777...62I} {777, 62}

\bibitem[\protect\citeauthoryear{{Kaneko}, {Preece}  \& et al.}{{Kaneko}
  et~al.}{2006}]{Kaneko2006}
{Kaneko} Y.,  {Preece} R.~D.,   et al. 2006, \mn@doi [ApJ] {10.1086/505911},
  \href {http://adsabs.harvard.edu/abs/2006ApJS..166..298K} {166, 298}

\bibitem[\protect\citeauthoryear{{Larsson}, {Racusin}  \& {Burgess}}{{Larsson}
  et~al.}{2015}]{Larsson2015}
{Larsson} J.,  {Racusin} J.~L.,   {Burgess} J.~M.,  2015, \mn@doi [ApJL]
  {10.1088/2041-8205/800/2/L34}, \href
  {http://adsabs.harvard.edu/abs/2015ApJ...800L..34L} {800, L34}

\bibitem[\protect\citeauthoryear{{Li}}{{Li}}{2018}]{Li2018}
{Li} L.,  2018, arXiv e-prints, \href
  {http://adsabs.harvard.edu/abs/2018arXiv181003129L} {}

\bibitem[\protect\citeauthoryear{{Lloyd} \& {Petrosian}}{{Lloyd} \&
  {Petrosian}}{2000}]{Lloyd2000}
{Lloyd} N.~M.,  {Petrosian} V.,  2000, \mn@doi [\apj] {10.1086/317125}, \href
  {http://adsabs.harvard.edu/abs/2000ApJ...543..722L} {543, 722}

\bibitem[\protect\citeauthoryear{{Lundman}, {Pe'er}  \& {Ryde}}{{Lundman}
  et~al.}{2013}]{Lundman2013}
{Lundman} C.,  {Pe'er} A.,   {Ryde} F.,  2013, \mn@doi [MNRAS]
  {10.1093/mnras/sts219}, \href
  {http://adsabs.harvard.edu/abs/2013MNRAS.428.2430L} {428, 2430}

\bibitem[\protect\citeauthoryear{{Medvedev}}{{Medvedev}}{2000}]{Medvedev2000}
{Medvedev} M.~V.,  2000, \mn@doi [\apj] {10.1086/309374}, \href
  {http://adsabs.harvard.edu/abs/2000ApJ...540..704M} {540, 704}

\bibitem[\protect\citeauthoryear{{Meegan} et~al.,}{{Meegan}
  et~al.}{2009}]{Meegan2009_GBM}
{Meegan} C.,  et~al., 2009, \mn@doi [\apj] {10.1088/0004-637X/702/1/791}, \href
  {http://adsabs.harvard.edu/abs/2009ApJ...702..791M} {702, 791}

\bibitem[\protect\citeauthoryear{{M{\'e}sz{\'a}ros}, {Ramirez-Ruiz}  \& et
  al.}{{M{\'e}sz{\'a}ros} et~al.}{2002}]{Meszaros2002}
{M{\'e}sz{\'a}ros} P.,  {Ramirez-Ruiz} E.,   et al. 2002, \mn@doi [ApJ]
  {10.1086/342611}, \href {http://adsabs.harvard.edu/abs/2002ApJ...578..812M}
  {578, 812}

\bibitem[\protect\citeauthoryear{{Oganesyan}, {Nava}, {Ghirlanda}, {Melandri}
  \& {Celotti}}{{Oganesyan} et~al.}{2019}]{Oganesyan2019}
{Oganesyan} G.,  {Nava} L.,  {Ghirlanda} G.,  {Melandri} A.,   {Celotti} A.,
  2019, arXiv e-prints, \href
  {http://adsabs.harvard.edu/abs/2019arXiv190411086O} {}

\bibitem[\protect\citeauthoryear{{Paczy{\'n}ski}}{{Paczy{\'n}ski}}{1986}]{Paczynski1986}
{Paczy{\'n}ski} B.,  1986, \mn@doi [ApJ] {10.1086/184740}, \href
  {http://adsabs.harvard.edu/abs/1986ApJ...308L..43P} {308, L43}

\bibitem[\protect\citeauthoryear{{Pe'er}}{{Pe'er}}{2008}]{Peer2008}
{Pe'er} A.,  2008, \mn@doi [\apj] {10.1086/588136}, \href
  {http://adsabs.harvard.edu/abs/2008ApJ...682..463P} {682, 463}

\bibitem[\protect\citeauthoryear{{Pe'er}, {M{\'e}sz{\'a}ros}  \&
  {Rees}}{{Pe'er} et~al.}{2006}]{Peer2006}
{Pe'er} A.,  {M{\'e}sz{\'a}ros} P.,   {Rees} M.~J.,  2006, \mn@doi [ApJ]
  {10.1086/501424}, \href {http://adsabs.harvard.edu/abs/2006ApJ...642..995P}
  {642, 995}

\bibitem[\protect\citeauthoryear{{Petrosian}}{{Petrosian}}{1981}]{Petrosian1981}
{Petrosian} V.,  1981, \mn@doi [\apj] {10.1086/159517}, \href
  {http://adsabs.harvard.edu/abs/1981ApJ...251..727P} {251, 727}

\bibitem[\protect\citeauthoryear{{Preece}, {Briggs}, {Mallozzi}, {Pendleton},
  {Paciesas}  \& {Band}}{{Preece} et~al.}{1998}]{Preece1998}
{Preece} R.~D.,  {Briggs} M.~S.,  {Mallozzi} R.~S.,  {Pendleton} G.~N.,
  {Paciesas} W.~S.,   {Band} D.~L.,  1998, \mn@doi [ApJL] {10.1086/311644},
  \href {http://adsabs.harvard.edu/abs/1998ApJ...506L..23P} {506, L23}

\bibitem[\protect\citeauthoryear{{Rees} \& {M{\'e}sz{\'a}ros}}{{Rees} \&
  {M{\'e}sz{\'a}ros}}{1994}]{Rees1994}
{Rees} M.~J.,  {M{\'e}sz{\'a}ros} P.,  1994, \mn@doi [ApJL] {10.1086/187446},
  \href {http://adsabs.harvard.edu/abs/1994ApJ...430L..93R} {430, L93}

\bibitem[\protect\citeauthoryear{{Rees} \& {M{\'e}sz{\'a}ros}}{{Rees} \&
  {M{\'e}sz{\'a}ros}}{2005}]{Rees&Meszaros2005}
{Rees} M.~J.,  {M{\'e}sz{\'a}ros} P.,  2005, \mn@doi [ApJ] {10.1086/430818},
  \href {http://adsabs.harvard.edu/abs/2005ApJ...628..847R} {628, 847}

\bibitem[\protect\citeauthoryear{{Rybicki} \& {Lightman}}{{Rybicki} \&
  {Lightman}}{1986}]{RybickiLightman}
{Rybicki} G.~B.,  {Lightman} A.~P.,  1986, {Radiative Processes in
  Astrophysics}

\bibitem[\protect\citeauthoryear{{Ryde}}{{Ryde}}{2004}]{Ryde2004}
{Ryde} F.,  2004, \mn@doi [ApJ] {10.1086/423782}, \href
  {http://adsabs.harvard.edu/abs/2004ApJ...614..827R} {614, 827}

\bibitem[\protect\citeauthoryear{{Ryde}}{{Ryde}}{2005}]{Ryde2005}
{Ryde} F.,  2005, \mn@doi [ApJ] {10.1086/431239}, \href
  {http://adsabs.harvard.edu/abs/2005ApJ...625L..95R} {625, L95}

\bibitem[\protect\citeauthoryear{{Ryde}, {Axelsson}  \& et al.}{{Ryde}
  et~al.}{2010}]{Ryde2010}
{Ryde} F.,  {Axelsson} M.,   et al. 2010, \mn@doi [ApJL]
  {10.1088/2041-8205/709/2/L172}, \href
  {http://adsabs.harvard.edu/abs/2010ApJ...709L.172R} {709, L172}

\bibitem[\protect\citeauthoryear{{Ryde}, {Pe'er}  \& et al.}{{Ryde}
  et~al.}{2011}]{Ryde2011}
{Ryde} F.,  {Pe'er} A.,   et al. 2011, \mn@doi [MNRAS]
  {10.1111/j.1365-2966.2011.18985.x}, \href
  {http://adsabs.harvard.edu/abs/2011MNRAS.415.3693R} {415, 3693}

\bibitem[\protect\citeauthoryear{{Ryde}, {Lundman}  \& {Acuner}}{{Ryde}
  et~al.}{2017}]{Ryde2017}
{Ryde} F.,  {Lundman} C.,   {Acuner} Z.,  2017, \mn@doi [\mnras]
  {10.1093/mnras/stx2019}, \href
  {http://adsabs.harvard.edu/abs/2017MNRAS.472.1897R} {472, 1897}

\bibitem[\protect\citeauthoryear{{Ryde}, {Yu}, {Dereli-B{\'e}gu{\'e}},
  {Lundman}, {Pe'er}  \& {Li}}{{Ryde} et~al.}{2019}]{Ryde2019}
{Ryde} F.,  {Yu} H.-F.,  {Dereli-B{\'e}gu{\'e}} H.,  {Lundman} C.,  {Pe'er} A.,
    {Li} L.,  2019, \mn@doi [\mnras] {10.1093/mnras/stz083}, \href
  {http://adsabs.harvard.edu/abs/2019MNRAS.tmp...65R} {}

\bibitem[\protect\citeauthoryear{{Sakamoto} et~al.,}{{Sakamoto}
  et~al.}{2009}]{Sakamoto2009}
{Sakamoto} T.,  et~al., 2009, \mn@doi [\apj] {10.1088/0004-637X/693/1/922},
  \href {http://adsabs.harvard.edu/abs/2009ApJ...693..922S} {693, 922}

\bibitem[\protect\citeauthoryear{{Scargle}, {Norris}, {Jackson}  \&
  {Chiang}}{{Scargle} et~al.}{2013}]{Scargle2013}
{Scargle} J.~D.,  {Norris} J.~P.,  {Jackson} B.,   {Chiang} J.,  2013, \mn@doi
  [\apj] {10.1088/0004-637X/764/2/167}, \href
  {http://adsabs.harvard.edu/abs/2013ApJ...764..167S} {764, 167}

\bibitem[\protect\citeauthoryear{{Stern} \& {Poutanen}}{{Stern} \&
  {Poutanen}}{2004}]{SternPoutanen2004}
{Stern} B.~E.,  {Poutanen} J.,  2004, \mn@doi [\mnras]
  {10.1111/j.1365-2966.2004.08163.x}, \href
  {http://adsabs.harvard.edu/abs/2004MNRAS.352L..35S} {352, L35}

\bibitem[\protect\citeauthoryear{{Tavani}}{{Tavani}}{1996}]{Tavani1996}
{Tavani} M.,  1996, \mn@doi [ApJ] {10.1086/177551}, \href
  {http://adsabs.harvard.edu/abs/1996ApJ...466..768T} {466, 768}

\bibitem[\protect\citeauthoryear{{Vianello}}{{Vianello}}{2018}]{Vianello2018}
{Vianello} G.,  2018, \mn@doi [\apjs] {10.3847/1538-4365/aab780}, \href
  {http://adsabs.harvard.edu/abs/2018ApJS..236...17V} {236, 17}

\bibitem[\protect\citeauthoryear{{Vianello}, {Gill}, {Granot}, {Omodei},
  {Cohen-Tanugi}  \& {Longo}}{{Vianello} et~al.}{2017}]{Vianello2017}
{Vianello} G.,  {Gill} R.,  {Granot} J.,  {Omodei} N.,  {Cohen-Tanugi} J.,
  {Longo} F.,  2017, preprint, \href
  {http://adsabs.harvard.edu/abs/2017arXiv170601481V} {} (\mn@eprint {arXiv}
  {1706.01481})

\bibitem[\protect\citeauthoryear{{Vianello}, {Gill}, {Granot}, {Omodei},
  {Cohen-Tanugi}  \& {Longo}}{{Vianello} et~al.}{2018}]{VianelloGill2018}
{Vianello} G.,  {Gill} R.,  {Granot} J.,  {Omodei} N.,  {Cohen-Tanugi} J.,
  {Longo} F.,  2018, \mn@doi [\apj] {10.3847/1538-4357/aad6ea}, \href
  {http://adsabs.harvard.edu/abs/2018ApJ...864..163V} {864, 163}

\bibitem[\protect\citeauthoryear{{Vurm}, {Beloborodov}  \& {Poutanen}}{{Vurm}
  et~al.}{2011}]{Vurm2011}
{Vurm} I.,  {Beloborodov} A.~M.,   {Poutanen} J.,  2011, \mn@doi [ApJ]
  {10.1088/0004-637X/738/1/77}, \href
  {http://adsabs.harvard.edu/abs/2011ApJ...738...77V} {738, 77}

\bibitem[\protect\citeauthoryear{{Vurm}, {Lyubarsky}  \& {Piran}}{{Vurm}
  et~al.}{2013}]{Vurm2013}
{Vurm} I.,  {Lyubarsky} Y.,   {Piran} T.,  2013, \mn@doi [ApJ]
  {10.1088/0004-637X/764/2/143}, \href
  {http://adsabs.harvard.edu/abs/2013ApJ...764..143V} {764, 143}

\bibitem[\protect\citeauthoryear{{Wang}, {Li}, {Moradi}  \& {Ruffini}}{{Wang}
  et~al.}{2019}]{Wang2019}
{Wang} Y.,  {Li} L.,  {Moradi} R.,   {Ruffini} R.,  2019, arXiv e-prints, \href
  {http://adsabs.harvard.edu/abs/2019arXiv190107505W} {}

\bibitem[\protect\citeauthoryear{{Yu} et~al.,}{{Yu} et~al.}{2016}]{Yu2016}
{Yu} H.-F.,  et~al., 2016, \mn@doi [\aap] {10.1051/0004-6361/201527509}, \href
  {http://adsabs.harvard.edu/abs/2016A%26A...588A.135Y} {588, A135}

\bibitem[\protect\citeauthoryear{{Yu}, {Dereli-B{\'e}gu{\'e}}  \& {Ryde}}{{Yu}
  et~al.}{2018}]{Yu2018}
{Yu} H.-F.,  {Dereli-B{\'e}gu{\'e}} H.,   {Ryde} F.,  2018, arXiv e-prints,
  \href {http://adsabs.harvard.edu/abs/2018arXiv181007313Y} {}

\bibitem[\protect\citeauthoryear{{Yu}, {Dereli-B{\'e}gu{\'e}}  \& {Ryde}}{{Yu}
  et~al.}{2019}]{Yu2019}
{Yu} H.-F.,  {Dereli-B{\'e}gu{\'e}} H.,   {Ryde} F.,  2019, arXiv e-prints,
  \href {http://adsabs.harvard.edu/abs/2018arXiv181007313Y} {}

\bibitem[\protect\citeauthoryear{{Zhang} \& {Yan}}{{Zhang} \&
  {Yan}}{2011}]{Zhang&Yan2011}
{Zhang} B.,  {Yan} H.,  2011, \mn@doi [ApJ] {10.1088/0004-637X/726/2/90}, \href
  {http://adsabs.harvard.edu/abs/2011ApJ...726...90Z} {726, 90}

\bibitem[\protect\citeauthoryear{{Zhang} et~al.,}{{Zhang}
  et~al.}{2018}]{ZhangBB2018}
{Zhang} B.-B.,  et~al., 2018, \mn@doi [Nature Astronomy]
  {10.1038/s41550-017-0309-8}, \href
  {http://adsabs.harvard.edu/abs/2018NatAs...2...69Z} {2, 69}

\makeatother
\end{thebibliography}






\bsp	
\appendix

\section{Bayesian formalism}

Bayesian inference applies the Bayes' theorem to infer and update parameter estimations, probabilities and distributions regarding a model after the experimental data is obtained. Bayes' theorem is given as

\begin{equation}
 p(\theta \mid y) = \frac{p(\theta) \, p(y \mid \theta)}{p(y)} 
\end{equation}
\noindent
where $p(\theta\mid y)$ is the posterior distribution of the estimated parameter $\theta$ given the data $y$. 

A model's ability of describing the data at hand can be put to the test by assessing the predictions it makes about the process that has created the observed data (y). If $y_{rep}$ is the replicated data set from the model's fit to the data, then the distribution of $y_{rep}$ conditional on y is called the $posterior$ $predictive$ $ distribution$ and is shown as \citep{bda3}, \\

\begin{equation}
p(y_{rep} \mid y) = \int_{}^{} p(y_{rep}, \theta \mid y) d\theta .
\end{equation}

A model is capable of describing the data gathered from a certain process if the future unknown observables ($\tilde{y}$) are successfully described by $y_{rep}$. In other words, the observed data $y$ and the replicated data $y_{rep}$ should be coming from the same generative process. In this paper, checks of posterior predictive distribution have been used to assess the quality of the fits done (see Appendix B).

\section{Checks of model fit viability}

\subsection{Corner plots and convergence diagnostics}
For all fits the corner plots displaying the posterior distributions of the resulting parameters have been visually checked. Posteriors which show clear bi or multi-modalities have been refit with refined initializations of the parameters to get unimodal normal-like posterior distributions for all simulations.

The convergence of the MCMC chains have also been assessed through trace plots to make sure that the true stationary distribution has been reached.\\

\subsection{Posterior Predictive Checks and QQ plots}
The viability of the spectral fits for the various simulations in this work has been assessed via posterior predictive checks (PPC). This includes the replication of the posterior distributions from the fitted spectra which are then compared to the observed data. A mismatch of the observed and replicated data would then indicate in what ways the model fails to represent the data, pointing to possible aspects of the model that should be modified or extended.

A compact way of comparing the observed and replicated data is via plotting the quantile-quantile (QQ) plots. Any significant deviation from the one to one line would then indicate the potential shortcomings of the current model to describe the data at hand \citep{Gelman2014}. 

Following the method of \citet{Burgess2018}, we have generated 500 realizations from each posterior given by the spectral fits. We then simulated the counts form these spectra and compared them to the observed data counts in QQ plots with 95\% and 68 \% quantiles superimposed. A deviation from 95\% quantile for a significant portion of the plot would indicate an unviable fit. We find no significant deviations in the presented simulation fits. We also compare the QQ plots of the fits to the cutoff powerlaw function to that of the real model used for the simulations. The former plot should not have great deviations from the latter for the fit to be acceptable. We find that this criterion is also satisfied in our fits.

\section{Original and simulated clusters comparisons}

\begin{figure}
 \includegraphics[width=\columnwidth]{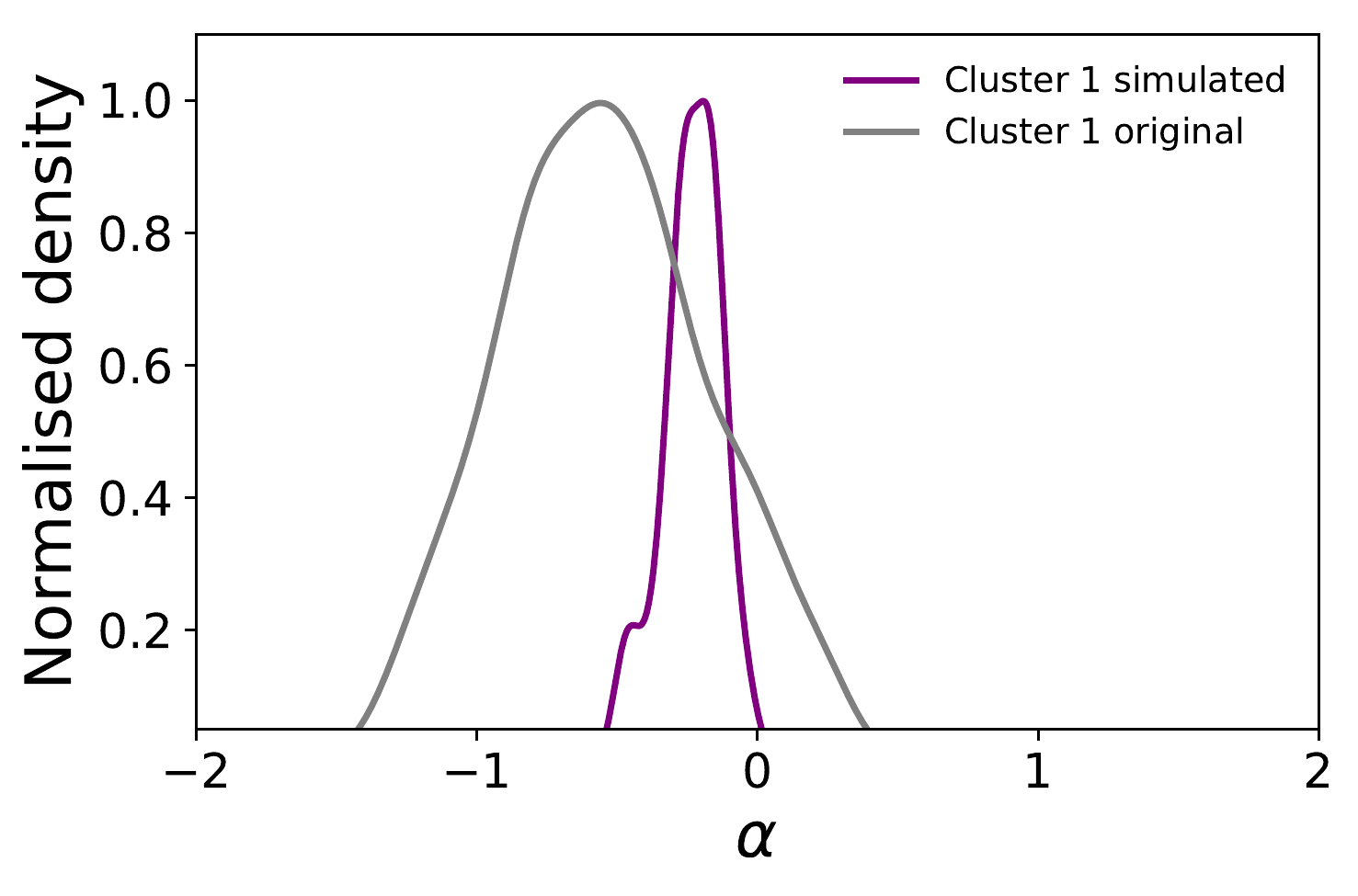}
 \caption{Distribution of $\alpha$-values for cluster 1 in \citet{Acuner2018}. The grey curve shown the KDE distribution for the GBM data, while the purple curve shows the KDE distribution of simulated non-dissipative photosphere spectra. The simulated spectra were assigned random $E_{\rm pk}$ values from the actual cluster distribution and hence have the same $E_{\rm pk}$-distributions.} 
 \label{fig:c1}
\end{figure}

\begin{figure}
 \includegraphics[width=\columnwidth]{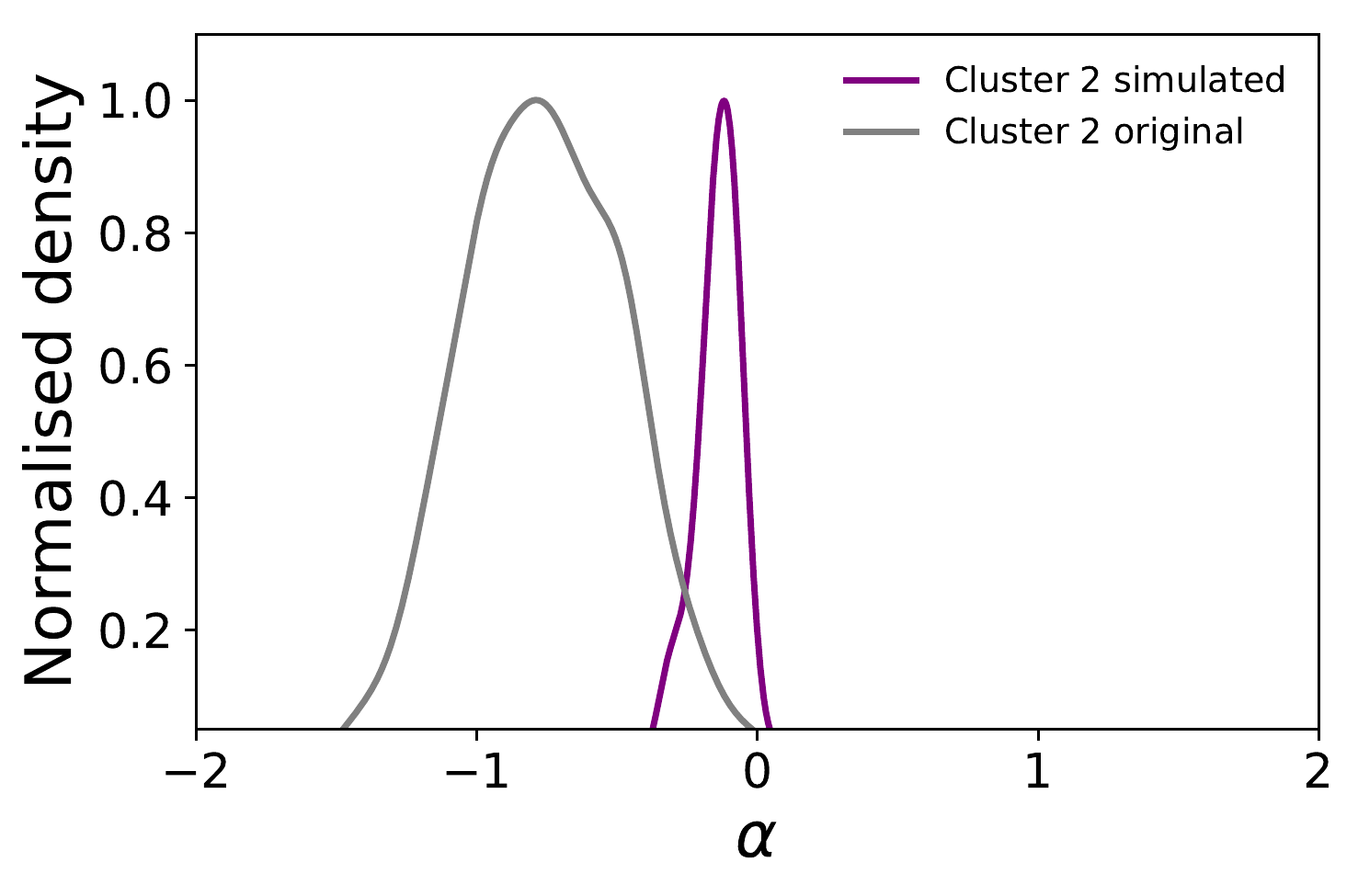}
 \caption{Same as Fig. \ref{fig:c1}, but for cluster 2 in \citet{Acuner2018}.
  }
 \label{fig:c2}
\end{figure}

\begin{figure}
 \includegraphics[width=\columnwidth]{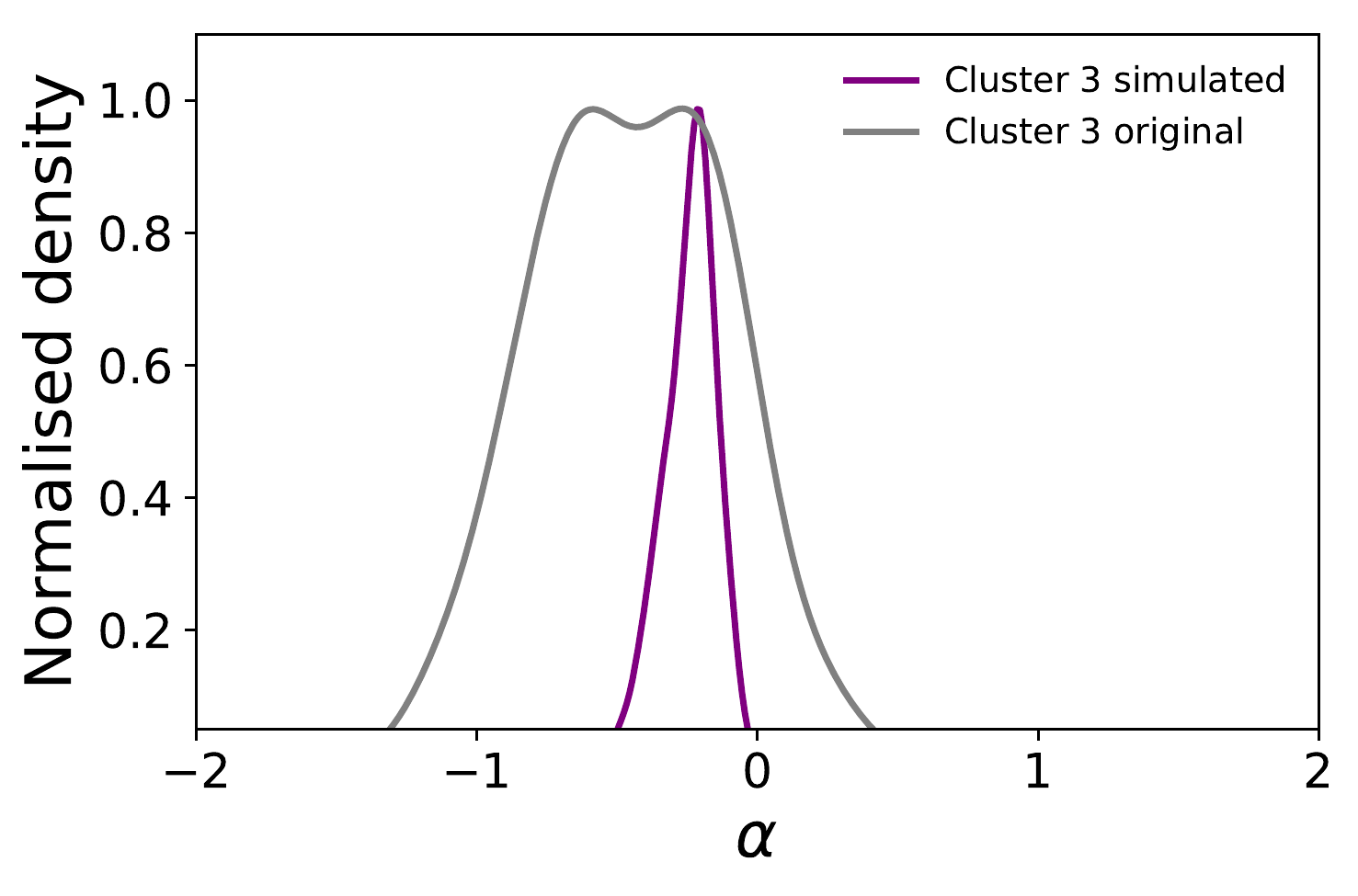}
 \caption{Same as Fig. \ref{fig:c1}, but for cluster 3 in \citet{Acuner2018}. These two distributions have the largest overlap.
  }
 \label{fig:c3}
\end{figure}

\begin{figure}
 \includegraphics[width=\columnwidth]{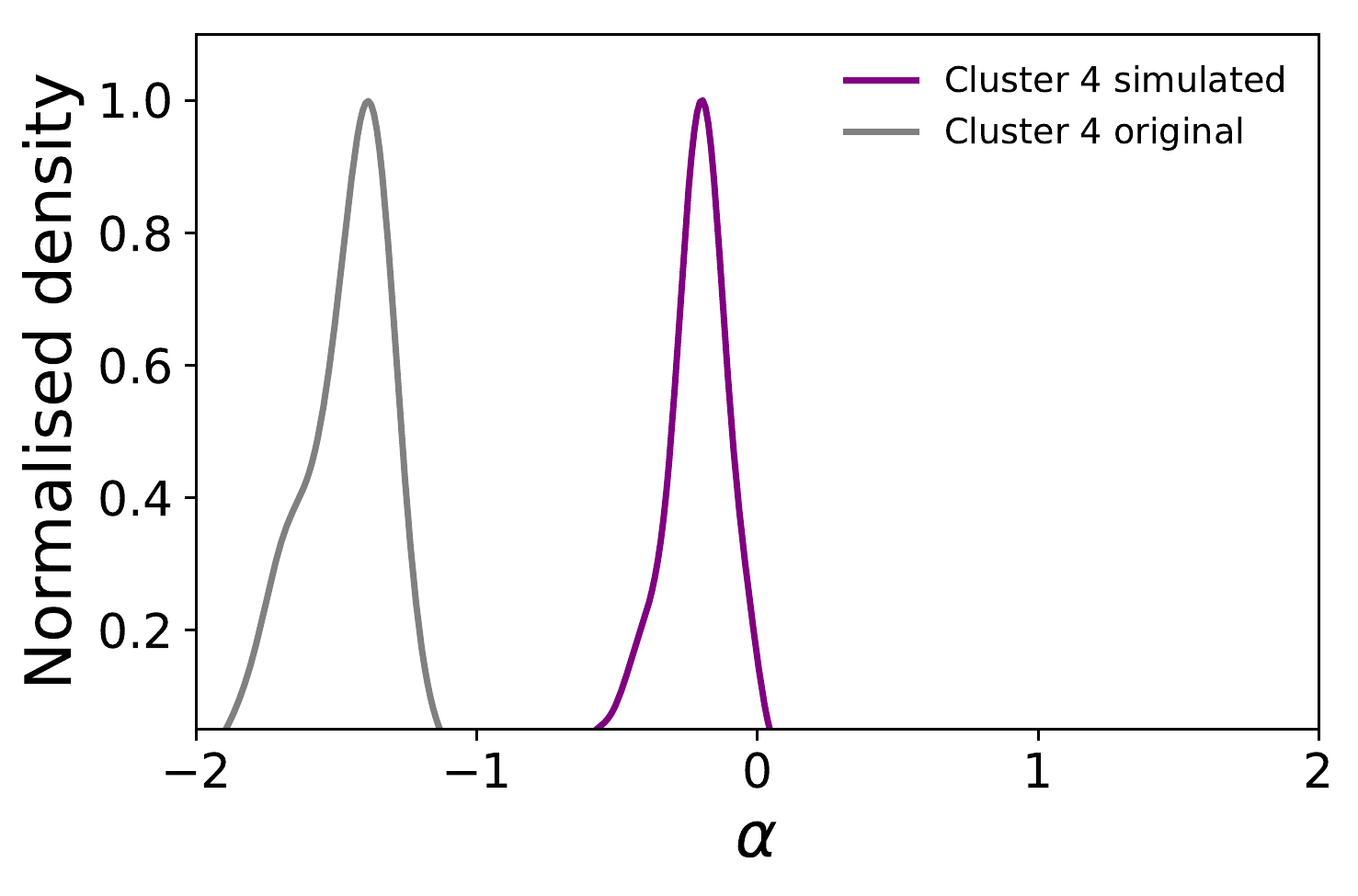}
 \caption{Same as Fig. \ref{fig:c1}, but for cluster 4 in \citet{Acuner2018}.
  }
 \label{fig:c4}
\end{figure}

\begin{figure}
 \includegraphics[width=\columnwidth]{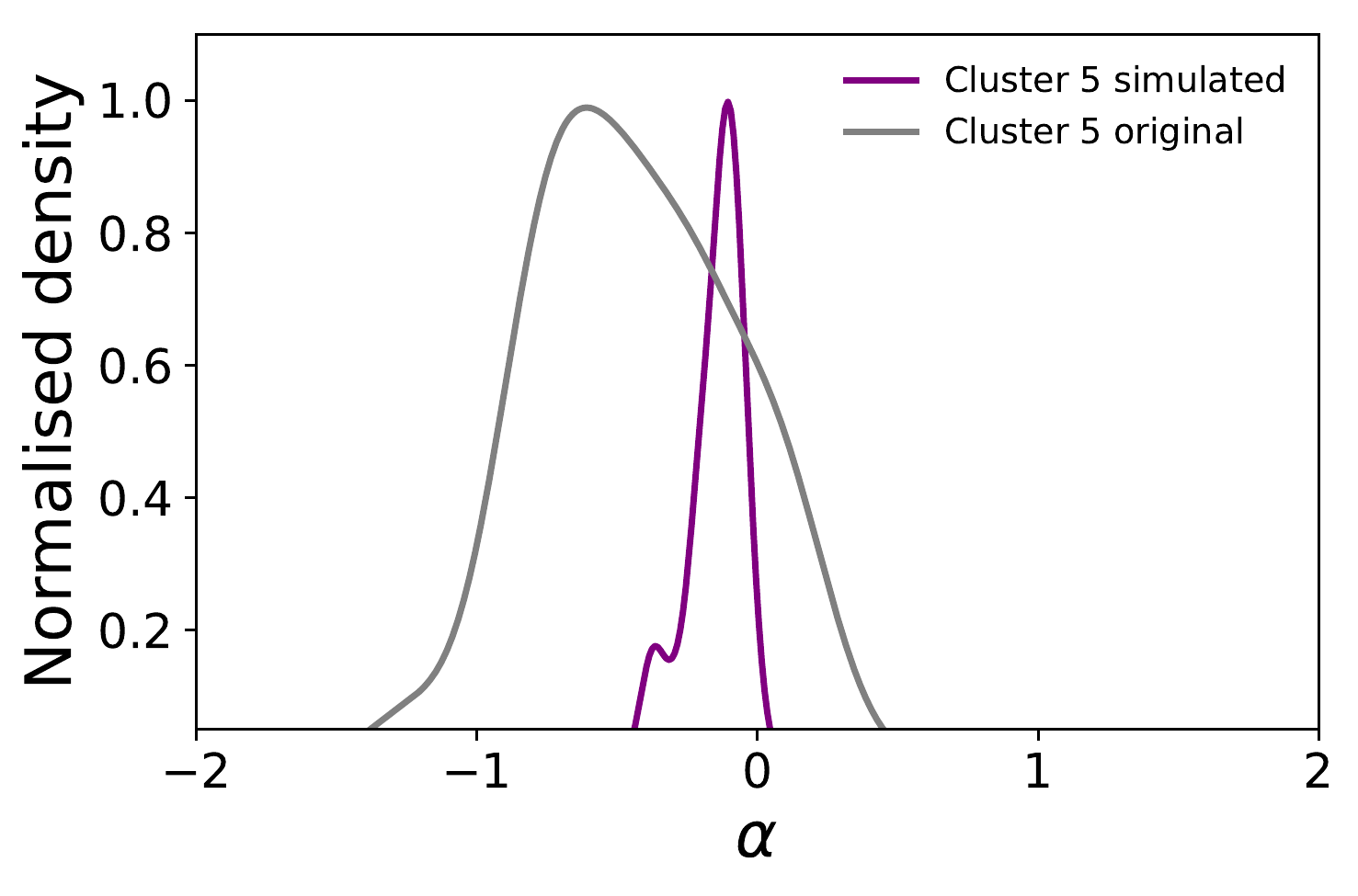}
 \caption{Same as Fig. \ref{fig:c1}, but for cluster 5 in \citet{Acuner2018}. 
  }
 \label{fig:c5}
\end{figure}

\label{lastpage}
\end{document}